# A Low-Complexity Diversity-Preserving Universal Bit-Flipping Enhanced Hard Decision Decoder for Arbitrary Linear Codes

**Praveen Sai Bere** *, *Student Member, IEEE*, **Mohammed Zafar Ali Khan** †, *Senior Member, IEEE* **and Lajos Hanzo** ‡, *Life fellow, IEEE*

*Department of Electrical Engineering, Indian Institute of Technology Hyderabad, Hyderabad, India,

ee19resch01001@iith.ac.in

†Department of Electrical Engineering, Indian Institute of Technology Hyderabad, Hyderabad, India, zafar@iith.ac.in

‡School of Electronics and Computer Science, University of Southampton, Southampton, United Kingdom,

lh@ecs.soton.ac.uk

**ABSTRACT** V2X (Vehicle-to-everything) communication relies on short messages for short-range transmissions over a fading wireless channel, yet requires high reliability and low latency.

Hard-decision decoding sacrifices the preservation of diversity order, leading to pronounced performance degradation in fading channels. By contrast, soft-decision decoding retains diversity order, albeit at the cost of increased computational complexity.

We introduce a novel enhanced hard-decision decoder termed as the Diversity Flip decoder (DFD) designed for preserving the diversity order. Moreover, it exhibits 'universal' applicability to all linear block codes. For a $\mathscr{C}(n,k)$ code having a minimum distance $d_{\min}$, the proposed decoder incurs a worst-case complexity order of $2^{(d_{\min}-1)} - 1$. Notably, for codes having low $d_{\min}$, this complexity represents a significant reduction compared to the popular soft and hard decision decoding algorithms. Due to its capability of maintaining diversity at a low complexity, it is eminently suitable for applications such as V2X (Vehicle-to-everything), IoT (Internet of Things), mMTC (Massive Machine type Communications), URLLC (Ultra-Reliable Low Latency Communications) and WBAN (Wireless Body Area Networks) for efficient decoding with favorable performance characteristics. The simulation results provided for various known codes and decoding algorithms validate the performance versus complexity benefits of the proposed decoder.

## I. Introduction

**T**HE adoption of Vehicle-to-everything (V2X) technology, which includes V2V (vehicle-to-vehicle), V2N (vehicle-to-network communications), V2I (vehicle-to-infrastructure) and V2P (vehicle-to-pedestrian) significantly enhances road safety, traffic flow efficiency, as well as the availability of entertainment and information services. ITS (Intelligent transportation systems) can be advanced through the implementation of V2X communications, facilitating cooperative operations [1]–[4].

The variable length of periodical broadcast message payloads exchanged between two User Equipment (UE) facil-itating V2X applications typically falls within the range of 50 to 300 bytes. In the case of event-triggered messages, the message size can extend to a maximum of 1200 bytes. The stringent latency requirements of V2V/V2P applications mandate a maximum end-to-end latency of 100 milliseconds between two UE regardless communicating directly or through a Roadside Unit (RSU). Similarly, for V2I applications, the maximum acceptable latency between a UE and an RSU is 100 milliseconds [5]–[7].

The 3GPP standardization body has established three principal service categories for 5G networks [8]. One of these categories is the enhanced mobile broadband (eMBB) mode, which caters to bandwidth-intensive applications such as high-resolution video streaming, augmented reality and virtual reality. Another category is mMTC (massive machine-type communication), which aims for accommodating a vast number of machine-type devices while ensuring optimal power efficiency for a prolonged battery recharge period. The third service category, URLLC (ultra-reliable and low latency communication), focuses on supporting time-sensitive applications and services requiring minimal delays.

Notably, mMTC and URLLC application services operate with shorter block lengths, representing distinct segments within these service categories.

The emergence of 5G NR technology has opened up new possibilities for Machine-Type Communication (MTC) services, which are anticipated to play a pivotal role in the Internet of Things (IoT) [9], especially in the context of the Industrial Internet of Things (IIoT) [10]. These MTC services can be categorized into two main types: mMTC (massive machine-type communications) [11] and URLLC (ultra-reliable low-latency communications) [12]. These categories delineate the diverse nature of MTC applications, catering for high-volume connectivity in the case of mMTC and stringent reliability as well as low latency requirements in the case of URLLC [12]–[14]. Short block lengths are a common feature in the realm of both mMTC and URLLC services.

mMTC applications, operating with short block lengths, necessitate low power consumption to prolong the battery recharge period, considering that a majority of these devices will operate on battery power. The need for an efficient decoder arises for mMTC applications relying on short block lengths. By contrast, URLLC must achieve both ultra-low latency and ultra-high reliability, which often pose conflicting demands. Nevertheless, to approach Shannon's capacity, legacy systems predominantly relied on coding schemes that encode information bits into longer codewords except for short control messages. But using longer codewords increases the latency jeopardizing the requirement of URLLC [15].

Maximum-minimum distance codes like BCH and Reed-Solomon codes developed back in the 1960s, have found renewed significance in the context of URLLC applications [16]–[20]. These codes enable the assignment of short codewords to information words, but require theoretically shortened mother codes for creating diverse code rates. On the other hand, RLCs (Random linear codes) offer flexibility in terms of their code rates but pose computational challenges [21], [22]. However, conventional hard-decision decoders or algebraic decoders of these short codes exhibit inadequate performance in practical scenarios involving fading channels due to their inability to maintain full diversity order[1] of $d_{min}$

. Soft-decision maximum-likelihood decoding (MLD) has the potential to preserve diversity over fading channels. However, its computational complexity is increased, rendering it impractical for light-weight real-world applications. Furthermore, the provision of soft-information may also be limited due to the computational complexity associated with soft-output detectors. Additionally, the iterative exchange of soft information between the detector and decoder tends to increase the processing latency. Therefore, an essential demand emerges for conceiving an efficient hard-decision decoder that guarantees diversity preservation in fading channels at a low complexity and low latency.

Numerous decoding techniques have been proposed in the literature, as indicated by the references [25]–[42]. Within this context, Information Set Decoding (ISD) is a notable topic, as evident from references [43]–[49]. Bossert *et al.* [50] specifically explore the application of ISD to BCH codes for transmission over a Binary Symmetric Channel (BSC). However, it is essential to highlight that many of these decoding techniques exhibit high computational complexity and do not guarantee the preservation of diversity in fading channels. On the other hand, the soft Maximum Likelihood Decoding (MLD) of Conway and Sloane [51] maintains diversity, but suffers from a high computational complexity, which grows exponentially with the message length $k$.

As a further advance, Chase introduced a class of algorithms that preserve diversity [52]. However, these algorithms depend on algebraic decoders and cannot operate as stand-alone universal decoders.

The recently introduced hard-decision GRAND (Guessing Random Additive Noise Decoding) solution of Duffy and Medard [53], [54] allows for universal decoding of linear codes. The GRAND technique is particularly efficient for high-rate codes with short block lengths [53]. However, being a hard decision decoder, the GRAND approach fails to maintain diversity over fading channels. Numerous soft variants of GRAND have been introduced in the literature, as detailed in [55]–[59]. Abbas *et al.* proposed Fading-GRAND [60] for fading channels. In the same scenario, GRAND with pseudo-soft information [61] is proposed which relies on pseudo-soft information. Chatzigeorgiou *et al.* proposed Symbol level GRAND in [22], [62], while [63] recommends employing Random Linear Codes (RLCs) along with GRAND for Ultra-Reliable Low Latency Communication (URLLC) over massive Multiple Input Multiple Output (mMIMO) systems. Given the inherent high complexity of the GRAND technique [64], the Hybrid Metric-based GRAND (HMGRAND) has been introduced as a viable alternative for Cyclic Redundancy Check (CRC) codes commonly employed in IoT applications [64]. This method utilizes an algebraic decoder specifically designed for CRC

---

[1]The diversity order [23] is defined as

$$\text{Diversity order} = \lim_{SNR \to \infty} \frac{-\log[P_e(SNR)]}{\log[SNR]}$$

The maximum diversity order that is obtainable by soft-decision decoding for a code $\mathcal{C}(n,k)$ with minimum distance $d_{min}$ in a fading channel is $d_{min}$ [24].

codes, making it suitable for IoT applications. However, it is worth noting that these decoders do not provide any insight into diversity preservation.

Patel *et al.* [24], [65] introduced flip decoding (FD) as a powerful technique for Parity Check (PC) codes also known as Single Parity Check codes (SPC) [66]. FD has been shown to attain a second-order diversity for parity check codes, akin to that derived from soft decision decoding. Notably, FD exhibits a decoding complexity that scales linearly with the block length $n$. This breakthrough allows PC codes, traditionally used for error detection, to be effectively harnessed for error correction in Rayleigh fading channels. The potential of FD in achieving dual-radio diversity using PC codes was explored by Ahmed in [67], wherein PC codes are applied above the MAC layer and decoding is also carried out above the MAC layer to glean the benefits of both offloading and duplication. The employment of PC codes at the MAC layer presents a solution to the challenge of combing offloading and duplication for multi-radios [68]–[81] for dual radios. Note that soft decoding is cumbersome above the MAC layer [82] and hence extending the coding and decoding above the MAC layer for more than two radios has certain requirements that are listed below:

1) Low complexity encoding: This requirement has restricted the use of codes to cyclic codes (including CRC codes) that have efficient low-complexity implementation using shift registers.

2) Low complexity decoding limits the decoding to hard decision techniques, since passing the soft values to the MAC or to the layers above the MAC is practically impossible.

3) Availability of RSSI: Since, all standards pass RSSI values to the MAC layer, this could be used for enhancing the hard decision decoding.

Independently, the use of flip decoding was further expanded to Hamming codes in [83] and CRC codes in [84]. These extensions markedly enhance the error-correction capability of Hamming and CRC codes in contrast to conventional hard-decision decoding approaches, all the while preserving the diversity order. Notably, the decoding complexity for both code types retains linearity in $n$.

Motivated by these benefits, we extend flip decoding to a broad family of linear codes, rendering it a 'universal' decoder that can be employed for any linear block code. The proposed decoder, which we have renamed as diversity flip decoder (DFD) to highlight its diversity preserving nature, attains an identical diversity order ($d_{\min}$) to soft-decision decoding despite its low complexity. This makes the proposed diversity flip decoder especially well-suited for decoding in fading channels, where diversity plays a pivotal role in alleviating the impact of fading.

With a particular emphasis on applications within wireless body area networks (WBANs), the IEEE 802.15.6 standard is designed to facilitate short-range wireless communication [85]. To ensure the reliability of communication, this standard incorporates $(63, 51)$ BCH codes and $(31, 19)$ shortened BCH codes, which are specifically tailored for WBAN applications. Consequently, researchers have concentrated their efforts on devising efficient decoders for BCH codes in the context of WBAN applications, as evidenced in [86]–[88]. However, these existing decoders predominantly rely on the BCH decoding algorithm, notably the Peterson algorithm [89]. By contrast, the proposed DFD technique outperforms them, despite its significantly lower complexity.

## A. Our Main Contributions
Our novel contributions are boldly contrasted to the literature in Table I and further detailed as follows:

- **Universal Applicability**: We introduce a universal DFD technique designed to preserve diversity in fading channels for all linear block codes.

- **Enhanced Error Correction:** We demonstrate that the proposed DFD significantly improves the error correction capabilities of linear block codes. Specifically, it can correct up to $(d_{\min}-1)$ errors, but not all $(d_{\min}-1)$ errors.

- **Preserved Diversity Order:** We show that our proposed decoder achieves a diversity order equal to '$d_{\min}$', aligning it with the diversity order of soft-decision decoding techniques, which is a critical aspect for improving reliability in fading channels.

- **Low Complexity:** The inherent advantage of the proposed DFD technique is its low computational complexity, when compared to well-established soft-decision and hard-decision decoders. This paves the way for its practical implementation.

- **Theoretical and Empirical Validation:** We provide comprehensive theoretical analyses complemented by our simulations to confirm its diversity order.

- **Enabling coding/decoding at MAC and above layers:** The proposed enhanced hard-decision decoder is particularly suitable for Multi-radios to achieve the benefits of both offloading and duplication.

- **Application Versatility:** The above benefits position the proposed DFD technique as an ideal solution for short-block length applications in V2X, WBAN, IoT, mMTC, and URLLC scenarios, where low complexity and high performance are paramount.

In Table I, the Chase decoder, OSD, ISD, ORBGRAND offer a correction capability higher than $t$ ($= \left\lfloor \frac{d_{\min}-1}{2} \right\rfloor$) since they utilize soft information. By contrast, GRAND is capable of correcting up to a maximum of $t$ errors. The OSD, ISD, GRAND, and ORBGRAND decoders are considered universal, as they are applicable to all types of codes. However, Chase decoders lack universality, since they are unsuitable for codes like Parity check codes or CRC codes due to their reliance on specific hard-decision decoders. Consequently, Chase decoders cannot function as standalone decoders, unlike the OSD, ISD, GRAND, and ORBGRAND decoders. Chase decoders maintain diversity over fading channels, whereas GRAND does not preserve the diversity order under fading conditions. For OSD, ISD, and ORBGRAND, the preservation of diversity order is not guaranteed at all times.

### B. Structure of the Paper

In Section II, we commence with an overview of the system setup, followed by the relevant concepts. The state-of-the-art is briefly summarized in Section C. Section III introduces our diversity-preserving DFD technique, while Section IV presents the derivation of an upper bound on the probability of error and the diversity order of DFD. Section V provides our simulation results. In Subsection A, DFD's achievement of the diversity order $d_{min}$ is demonstrated. In Subsection B, the DFD's performance for different codeword lengths and code rates are presented, while Subsection C offers our performance versus complexity analysis. Section VI introduces the Extended Diversity Flip Decoder (EDFD), which represents a further enhancement in performance. Subsequently, the simulation results for the EDFD are detailed in the subsequent section, Section VII. Finally, our conclusions are drawn in Section VIII.

## II. Preliminaries

### A. System setup

Consider a binary linear block code $\mathscr{C}(n, k)$ having a parity check matrix (PCM) $\mathbf{H_{(n-k) \times n}}$ and a generator matrix (GM) $\mathbf{G_{k \times n}}$. Here, $n$ denotes the codeword length and $k$ represents the dimension of the code $\mathscr{C}(n, k)$, which possesses a minimum distance of $d_{min}$.

We adopt the following notation: a matrix is denoted by $\mathbf{X}$, while a vector is denoted by $\underline{y}$; a row vector with a length of $n$ is denoted by $\underline{y^n}$; $\underline{0^n}$ indicates a row vector of zeros with a length of $n$, and $y$ stands for a scalar; $w_H(\underline{y})$ denotes the Hamming weight of a binary vector $\underline{y}$; $\underline{q} \oplus \underline{p}$ formulates the binary XOR operation of the corresponding elements of the pair of binary vectors $\underline{q}, \underline{p}$, where we have:

$$\underline{q} \oplus \underline{p} = [q_1 \oplus p_1 \quad q_2 \oplus p_2 \quad q_3 \oplus p_3 \quad \cdots \quad q_n \oplus p_n]_{1 \times n}$$

with $q_j, p_j \; \epsilon \; \{0, 1\}$.

Let $\underline{c} = [c_1 \quad c_2 \quad c_3 \quad \cdots \quad c_n]_{1 \times n} \in \mathscr{C}(n, k)$ be a legitimate transmitted codeword. In the case of a practical scenario involving a Rayleigh fading channel, BPSK modulation is assumed for simplicity. Under BPSK modulation, $\underline{c}$ is mapped to a symbol sequence $\underline{s} = [s_1 \quad s_2 \quad s_3 \quad \cdots \quad s_n]_{1 \times n}$, where we have $s_j = (-1)^{c_j}$. Thus, $s_j \in \{-1, 1\}, \forall j$. Then, the received signal can be represented as:

$$y_j = h_j s_j + w_j, \; 1 \leq j \leq n, \; h_j \geq 0, \quad (1)$$

where $w_j$ represents statistically independent complex Gaussian random variables with zero mean and variance of $N_0$. The coefficients $h_j$ denote identically distributed, independent Rayleigh fading coefficients, indicating a fast-fading channel assumption. In the context of a slow-fading channel, the codewords are assumed to be interleaved ideally. Let $\underline{h} = [h_1 \quad h_2 \quad \cdots \quad h_n]_{1 \times n}$ denote the Channel State Information (CSI) vector. Additionally, it is assumed that the receiver has perfect knowledge of the CSI.

The received vector is denoted by $\underline{y} = [y_1 \quad y_2 \quad \cdots \quad y_n]_{1 \times n}$. Applying a hard decision to $\underline{y}$, the received vector is represented as $\underline{r} = [r_1 \quad r_2 \quad \cdots \quad r_n]_{1 \times n}$.

### B. Known Results of Diversity

The linear block code $\mathscr{C}(n, k)$ with minimum distance $d_{min}$ and error correcting capability $t$ exhibits a diversity order of $(t + 1) = \left\lfloor \frac{d_{min}-1}{2} \right\rfloor + 1$ under hard decision decoding, while soft decision decoding achieves a diversity order of $d_{min}$ [24].

While hard decision decoding of $\mathscr{C}(n, k)$ offers lower complexity than soft decision decoding, it sacrifices diversity order, leading to significant performance degradation in fading channels. On the other hand, soft decision maximum likelihood decoding delivers excellent performance, but involves a comparison of $2^k$ legitimate codewords, resulting in high complexity. As the value of $k$ increases, the complexity of soft decision ML decoding grows exponentially.

### C. Previous Relevant Work

In this subsection, we present a comprehensive overview of Chase, OSD, ISD (Information Set Decoding), along with recent research related to GRAND (Guessing Random Additive Noise Decoding) decoders. We classify them into "guessing" noise or codeword type.

**Definition 1:**
Guessing codeword decoders are the decoders that search over codewords.

**Definition 2:**
Guessing noise decoders are the decoders that search over different noise patterns.

In Table II, various decoders are categorized into two groups: 'Guessing noise type' and 'Guessing codeword type'.

#### 1) Flip Decoding for Parity check codes
The idea of Flip decoding for PC codes appeared first in [65]. This methodology entails flipping the bit linked with the minimum Channel State Information (CSI) value wherever a parity mismatch arises at the receiver. Through this Flip Decoder (FD) technique, PC codes have showcased a notable diversity order of 2 in fading channels.

TABLE I: Distinguishing Our New Contributions from the State-of-the-Art

| Attribute | Decoder type | | | | | | |
|---|---|---|---|---|---|---|---|
| | Chase [52] | OSD [35], [46] | ISD [18], [43], [44] | GRAND [53] | ORBGRAND [58], [60], [61] | Proposed DFD | Proposed EDFD |
| Enhanced Error correction ($> t$) | ✓ | ✓ | ✓ | | ✓ | ✓ | ✓ |
| Universal decoder | | ✓ | ✓ | ✓ | ✓ | ✓ | ✓ |
| Stand-alone decoder | | ✓ | ✓ | ✓ | ✓ | ✓ | ✓ |
| Preservation of diversity | ✓ | | | | | ✓ | ✓ |
| Low Complexity | | | | | | ✓ | ✓ |
| Suitability for decoding at MAC and above MAC layers | | | | ✓ | | ✓ | ✓ |

TABLE II: Categorizing the decoders

| Decoder | Type |
|---|---|
| Chase | Guessing codeword type |
| OSD | Guessing codeword type |
| ISD | Guessing codeword type |
| GRAND | Guessing noise type |
| ORBGRAND | Guessing noise type |
| Proposed DFD | Guessing noise type |
| Proposed EDFD | Guessing noise type |

### 2) Chase decoders

In 1972, Chase introduced a class of decoding algorithms that leverage channel measurement information [52]. Here's a concise overview of the Chase decoding algorithms: Start with the received sequence denoted as $\underline{y}$. Introduce a test pattern, denoted as $\underline{T}$, added to $\underline{y}$ to produce a new sequence, $\underline{y}'$. For each $\underline{y}'$, utilize a hard-decision binary decoder to derive a new sequence, $\underline{y}''$. Let $\underline{z}'$ represent the corresponding error pattern. Then, the effective error pattern is determined by

$$\underline{z}_T = \underline{T} \oplus \underline{z}'.$$

Ultimately, the effective error pattern ($\underline{z}_T$) having the lowest analog weight [52] is chosen as the definitive error pattern, and its associated codeword is designated as the decoded codeword.

Chase presented three algorithms characterized by a progressively diminishing count of error patterns ($\underline{T}$) within the specified error pattern set.

*Algorithm-1*: This algorithm encompasses all error patterns having a binary Hamming weight of $\lfloor \frac{d_{min}}{2} \rfloor$. Consequently, a total of $\binom{n}{\lfloor \frac{d_{min}}{2} \rfloor}$ error patterns are included in this approach, which can be substantial in size.

*Algorithm-2*: This approach involves error patterns ($\underline{T}$) where all combinations of 1's are positioned within the $\lfloor \frac{d_{min}}{2} \rfloor$ lowest channel measurement information values.

Consequently, $2^{\lfloor \frac{d_{min}}{2} \rfloor}$ test error patterns are taken into account in this method.

*Algorithm-3*: In this approach, the complexity is further reduced. It involves considering test patterns with a Hamming weight of $i$, where $i$ represents the number of 1's positioned in slots corresponding to the $i$ lowest channel measurement values. The values of $i$ encompass:

$$i = 0, 1, 3, \ldots d_{min} - 1, \quad \text{when } d_{min} \text{ is even.}$$
$$i = 0, 2, 4, \ldots d_{min} - 1, \quad \text{when } d_{min} \text{ is odd.}$$

Consequently, the total count of error patterns considered will be $\lfloor (\frac{d_{min}}{2}) + 1 \rfloor$, resulting in a significant reduction in the number of error patterns.

**Remark 1:**

It's noteworthy that Chase decoders aren't universally applicable since their decoding relies on the specific hard-decision decoding algorithm of the code.

### 3) OSD

Let $\underline{\alpha} = [\alpha_1 \quad \alpha_2 \quad \cdots\cdots \quad \alpha_n]$ be a reliable vector holding the reliabilities of all $n$ bits of the received codeword $\underline{r}$. In OSD, the reliabilities are ordered in descending order to get $\underline{\alpha}_{\text{sort}} = [\alpha_{(1)} \quad \alpha_{(2)} \quad \cdots\cdots \quad \alpha_{(n)}]$. The corresponding permutation function is $\lambda_1$. Then, $\lambda_1$ is applied to $\underline{r}$ to get $\underline{r}'$. The same $\lambda_1$ is applied to the columns of the generator matrix $G$ to get $G'$. Then, the first $k$ independent columns of $G'$ are identified and selected as the first $k$ columns of $G''$. The remaining columns of $G'$ will be the next $(n - k)$ columns of $G''$. Let this permutation function be $\lambda_2$. The matrix $G''$ is converted into a systematic form to get $G_{OSD}$. Then, $\lambda_2$ is applied to $\underline{r}'$ in order to get $\underline{r}''$. The first $k$ bits in $\underline{r}''$ are denoted by $\underline{r}''_{\text{M}}$ and the remaining bits are denoted by $\underline{r}''_{\text{L}}$. In OSD, Test Error Patterns (TEPs) of length $k$ bits are employed for reprocessing. For order-$i$ reprocessing, TEPs ($\underline{e}$) up to Hamming weight of $i$ are considered. The codeword is reprocessed as follows

$$\underline{\tilde{c}}_\mathbf{e} = (\underline{r}''_{\text{M}} \oplus \underline{e}) \cdot G_{OSD}.$$

Let $\underline{\tilde{\alpha}} = [\tilde{\alpha}_1 \quad \tilde{\alpha}_2 \quad \ldots \quad \tilde{\alpha}_n] = \lambda_2[\lambda_1(\underline{\alpha})]$. The weighted Hamming distance is defined as follows:

$$d^{(W)}(\underline{r}'', \underline{\tilde{c}}_e) \triangleq \sum_{\substack{1 \le i \le n \\ r_i'' \ne \tilde{c}_{e_i}}} \tilde{\alpha}_i.$$

In OSD, a particular codeword corresponding to the minimum weighted Hamming distance is chosen as an optimal codeword ($\underline{\tilde{c}}_{\mathbf{opt}}$). Then, the appropriate inverse permutations are applied to $\underline{\tilde{c}}_{\mathbf{opt}}$ to get the final decoded codeword ($\underline{\hat{c}}_{\mathbf{opt}}$).

$$\underline{\hat{c}}_{\mathbf{opt}} = \lambda_1^{-1}[\lambda_2^{-1}(\underline{\hat{c}}_{\mathbf{opt}})].$$

### 4) ISD

ISD [18], [43] uses specific bit positions $(i_0, i_1, i_2, ..., i_{k-1})$ to calculate a systematic basis[2] ($B = 0, 1, 2, ..., k-1$) and if these positions contain linearly dependent columns, it extends the search until a systematic basis is found. Using this basis, ISD then encodes data with received bits $r_0, r_1, r_2..., r_{k-1}$ as information. It is preventing decoding failure. ISD can also become a list decoder, flipping information bits and choosing the codeword having the smallest Hamming distance.

### 5) GRAND and GRANDAB

The GRAND [53] technique capitalizes on the observation that as the channel conditions improve, the set of potential error patterns capable of altering a legitimate transmitted codeword becomes smaller. This insight enables a shift in focus from searching for the most likely transmitted codeword within the codebook to estimating the most probable error pattern.

To determine whether an error pattern corresponds to a valid codeword, a code-book membership test is employed. GRAND organizes the prospective error patterns in descending order of likelihood and sequentially applies the membership test to each error pattern. The algorithm terminates and returns the first error pattern that successfully passes the membership test.

In contrast to GRAND, GRANDAB [59] not only halts at the first occurrence of an affirmative response but also terminates and indicates an erasure when the number of codebook queries exceeds a certain threshold known as Abandonment bit ($ABB$).

---

[2]systematic basis is the linearly independent $k$ columns. For example a systematic code $\mathcal{C}(n, k)$ with generator matrix $G_{k \times n}$ is considered. Let the set $P \subset \{0, 1, 2, ....n-1\}$ represent a set of $k$ distinct positions. If the submatrix $G_P$ comprising columns $j \in S$, achieves full rank $k$, $P$ is termed an information set. These positions uniquely determine the remaining $n - k$ positions in any codeword.

### 6) ORBGRAND

Consider the transmission of $\underline{c}$ over a binary-input AWGN channel, with $\underline{r}$ as the received vector. The codeword bits can be sorted in ascending order of reliability based on $\underline{r}$, and a vector $\underline{s}$ is used to record the reliability order of each bit. For an error pattern vector $\underline{z}$, the logistic weight $w_l$ is defined as $w_l = \sum_{i=0}^{n-1}(s_i \times z_i)$. It follows that $w_l(z^{k_1}) < w_l(z^{k_2})$ for $k_1 < k_2$. Smaller values of $w_l(z^k)$ indicate higher likelihoods of the corresponding error patterns.

### 7) Fading-GRAND

The fading-GRAND of [60] leverages channel state information to identify a set of bit indices denoted as $\mathcal{I}$. This set is constructed based on a given threshold value ($\Delta$). The expression for $\Delta$ is defined as $\Delta = m\left(\frac{E_b}{N_0}\right) + b$, where $m$ and $b$ represent the optimal values that yield superior performance for different codes, as specified in [60]. In the next step, Test Error Patterns (TEPs) are generated by ensuring that the positions of 1s in the TEPs exclude the positions specified in the set $\mathcal{I}$.

### 8) GRAND using pseudo-soft information

The method described in [61] introduces an approach where the received bits are arranged according to their pseudo-soft information, which incorporates both the channel state information (CSI) and the colored additive noise (CAN) statistics, rather than conventional soft information. Subsequently, the remaining steps of the GRAND technique are applied without any modifications.

## III. Proposed Universal Diversity Flip Decoder

This section presents the proposed Diversity Flip decoder (DFD) conceived for all linear block codes. The essence of the proposed DFD lies in its operational principle. Within fading channels, the fading envelope has a more pronounced influence on errors than the Additive White Gaussian Noise (AWGN). Consequently, the DFD leverages the knowledge of Channel State Information (CSI) to rectify the errors by selectively flipping the bits associated with the lowest CSI values. These bits are the ones most likely to be erroneous, enabling their targeted flipping within the corrupted codeword in order to restore the integrity of the transmitted codeword.

### A. The Diversity Flip Decoder

Consider the linear block code denoted as $\mathscr{C}(n, k)$, possessing a minimum distance of $d_{\min}$. Let $\underline{c} = [c_1 \quad c_2 \quad \cdots \quad c_n]_{1 \times n} \in \mathscr{C}$ represent the transmitted codeword corresponding to the information word $\underline{m} = [m_1 \quad m_2 \quad \cdots \quad m_k]_{1 \times k}$. After a hard-decision process at the receiver, the received word is denoted as $\underline{r} = [r_1 \quad r_2 \quad \cdots \quad r_n]_{1 \times n}$. Furthermore, the vector $\underline{h} = [h_1 \quad h_2 \quad \cdots \quad h_n]_{1 \times n}$ is defined as the collection of ab-solute values characterizing the channel state information (CSI).

In the flip decoding process, we sort the elements of $\underline{h}$ in ascending order (similar to OSD [90]), resulting in the sorted vector denoted as $\underline{h}_{\mathbf{sort}} = [h_{(1)} \quad h_{(2)} \quad \cdots \quad h_{(n)}]_{1 \times n}$. Let '$\pi$' represent the corresponding permutation function. Applying the same permutation function '$\pi$' to the vector $\underline{r}$ yields the vector $\underline{r}_{\mathbf{sort}} = [r_{(1)} \quad r_{(2)} \quad \cdots \quad r_{(n)}]_{1 \times n}$. Thus, we can express it as:

$$\underline{h}_{\mathbf{sort}} = \pi(\underline{h}),$$
$$\underline{r}_{\mathbf{sort}} = \pi(\underline{r}),$$
$$\Rightarrow \underline{r} = \pi^{-1}(\underline{r}_{\mathbf{sort}}).$$

Define

$$d = d_{\min} - 1. \tag{2}$$

Then the initial $d$ bits in $\underline{r}_{\mathbf{sort}}$ represent the least reliable bits (LRB), while the remaining bits are referred to as the most reliable bits (MRB). This can be expressed as follows:

$$\underline{r}_{\mathbf{lrb}} = [r_{(1)} \quad r_{(2)} \quad r_{(3)} \quad \cdots \quad r_{(d)}]_{1 \times d},$$
$$\underline{r}_{\mathbf{mrb}} = [r_{(d+1)} \quad r_{(d+2)} \quad r_{(d+3)} \quad \cdots \quad r_{(n)}]_{1 \times (n-d)}.$$

Consequently, we have:

$$\underline{r}_{\mathbf{sort}} = [\underline{r}_{\mathbf{lrb}} \quad \underline{r}_{\mathbf{mrb}}].$$

Subsequently, we embark on an exploration of all conceivable patterns of flipping the least reliable bits (LRB). Each flip vector ($\underline{f}_{\mathbf{pv}}^{\mathbf{n}}$) is represented as a row vector of the same size as the codeword $\underline{c}^{\mathbf{n}}$. We formally define the set of flip vectors as $\Phi$, denoted as:

$$\Phi = \left\{ \underline{f}_{\mathbf{pv}}^{\mathbf{n}} \;\middle|\; \underline{f}_{\mathbf{pv}}^{\mathbf{n}} = \left[ \underline{f}_{\mathbf{p}}^{\mathbf{d}(i)} \quad \underline{0}^{\mathbf{n-d}} \right] \right\} \quad (1 \leq i \leq 2^d - 1).$$

For a given decimal number $i$, the flip pattern vector, denoted as $\underline{f}\mathbf{p}^{\mathbf{d}(i)}$, is constructed such that the least significant bit (LSB) takes the first position when converting the decimal number $i$ to a binary vector of length $d$. Additionally, $\underline{f}_{\mathbf{pv}}^{\mathbf{n}}$ is referred to as the 'flip vector'. To illustrate this, consider the BCH(15,7,5) code where $i = 3$, $d = 4$, and $n = 15$. We have:

$$\underline{f}_{\mathbf{p}}^{\mathbf{d}(3)} = [1 \quad 1 \quad 0 \quad 0]$$
$$\underline{f}_{\mathbf{pv}}^{\mathbf{n}} = [\underline{f}_{\mathbf{p}}^{\mathbf{d}(3)} \quad \underline{0}^{\mathbf{11}}]$$
$$= [1 \quad 1 \quad 0 \quad 0 \quad 0 \quad 0 \quad 0 \quad 0 \quad 0 \quad 0 \quad 0 \quad 0 \quad 0 \quad 0 \quad 0].$$

Please be aware that for a given '$d = d_{\min} - 1$', the set of flip patterns remains fixed and does not change with the type of code, as long as the minimum distance '$d_{\min}$' is constant. The composition of the final set of flip vectors is determined by appending corresponding zero vectors based on the codeword length 'n'. Therefore, the creation of the flip vector set is straightforward, has low complexity, and does not necessitate the repeated generation of flip vectors for each code type employed.

Let $\Phi(l)$ denote the $l^{th}$ element of the set $\Phi$. The cardinality of $\Phi$ is $2^d - 1$. The Flip decoding technique operates as follows: Each flip vector is sequentially selected from the set $\Phi$, and it is added (XORed) to the sorted received vector, $\underline{r}_{\mathbf{sort}}$, resulting in $\underline{r}_{\mathbf{t}}$. Then, the inverse permutation function of $\underline{r}_{\mathbf{t}}$ is applied to obtain the resultant vector $\underline{r}_{\mathbf{o}}$. This can be expressed as:

$$\underline{r}_{\mathbf{t}} = \underline{r}_{\mathbf{sort}} \oplus \underline{f}_{\mathbf{pv}},$$
$$\underline{r}_{\mathbf{o}} = \pi^{-1}(\underline{r}_{\mathbf{t}}).$$

Subsequently, the syndrome vector ($\underline{S}_{\mathbf{y}}$) is computed for $\underline{r}_{\mathbf{o}}$ using the matrix $\mathbf{H}$ as $\underline{S}_{\mathbf{y}} = \underline{r}_{\mathbf{o}}\mathbf{H}^T$. If the resultant syndrome ($\underline{S}_{\mathbf{y}}$) is non-zero, the next flip vector is selected from the set $\Phi$ and added to the vector $\underline{r}_{\mathbf{sort}}$. This iterative process continues until a zero syndrome is obtained. Upon obtaining a zero syndrome, the resultant vector ($\underline{r}_{\mathbf{o}}$) represents the decoded codeword ($\underline{c}_{\mathbf{o}}^{\mathbf{opt}}$), thereby concluding the process. However, if none of the flip vectors $\underline{f}_{\mathbf{pv}}^{\mathbf{n}}$ in the set $\Phi$ yields a valid codeword, Algorithm-1 returns the received word ($\underline{r}$) as the decoded codeword $\underline{c}_{\mathbf{o}}^{\mathbf{opt}}$. Subsequently, the further decoded information word $\mathbf{m}_{\mathbf{o}}^{\mathbf{opt}}$ is derived by employing the codebook $\mathscr{C}$.

Algorithm-1 provides the pseudo-code of the FD applicable to all linear block codes. In this algorithm, the '$\sim$' operator represents the logical NOT operation. The function '$sum(\underline{S}_{\mathbf{y}})$' calculates and returns the sum of the elements in vector $\underline{S}_{\mathbf{y}}$. Additionally, the function '$sort(\underline{h})$' yields the vector $\underline{h}_{\mathbf{sort}}$, where the elements of $\underline{h}$ are arranged in ascending order. It also provides a vector, $\underline{\mathbf{Ix}}$ which serves as the permutation function containing the permuted indices corresponding to the elements of $\underline{h}$. The functionality of '$sort(\underline{h})$' can be better understood through the following example.

*If $\underline{h} = [0.82 \quad 1.3 \quad 1.08 \quad 0.09 \quad 0.43 \quad 1.8 \quad 0.32]$;*

$[\underline{h}_{\mathbf{sort}}, \underline{\mathbf{Ix}}] = sort(\underline{h}); \quad then,$

$\underline{h}_{\mathbf{sort}} = [0.09 \quad 0.32 \quad 0.43 \quad 0.82 \quad 1.08 \quad 1.3 \quad 1.8]; and$

$\underline{\mathbf{Ix}} = [4 \quad 7 \quad 5 \quad 1 \quad 3 \quad 2 \quad 6];$

$\underline{\mathbf{Ix}}(1) = 4; \quad \underline{\mathbf{Ix}}(3) = 5.$

**Example 1:**
The Flip decoding process is elucidated through the examination of the following illustrative example involving a BCH(15,7,5) code. Let the transmitted codeword be
$\underline{c} = [1 \quad 0 \quad 0 \quad 1 \quad 1 \quad 0 \quad 1 \quad 1 \quad 1 \quad 0 \quad 0 \quad 0 \quad 1 \quad 0]_{1 \times n}.$

Consider the received codeword after hard-decision, along with the associated Channel State Information (CSI) vector at the receiver, denoted by

$\underline{r} = [1 \quad 0 \quad 0 \quad 1 \quad 1 \quad 1 \quad 1 \quad 1 \quad 1 \quad 0 \quad 0 \quad 0 \quad 0 \quad 1 \quad 0]_{1 \times 15},$
$\underline{h} = [1.0869 \quad 0.7561 \quad 2.496 \quad 1.8351 \quad 0.416$
$\qquad 0.1256 \quad 0.9395 \quad 1.6002 \quad 0.4133 \quad 1.6239$
$\qquad 0.0854 \quad 1.1069 \quad 0.817 \quad 0.9698 \quad 1.5772]_{1 \times 15}.$

When the elements of $\underline{h}$ are arranged in ascending order, resulting in $\underline{h}_{\mathbf{sort}}$, along with the corresponding permutation function $\pi$ represented as a vector $\underline{\mathbf{Ix}}$, applying this permutation function $\pi$ to $\underline{\mathbf{r}}$ yields $\underline{\mathbf{r}}_{\mathbf{sort}}$.

$$[\underline{\mathbf{h}}_{\mathbf{sort}}, \underline{\mathbf{Ix}}] = sort(\underline{\mathbf{h}}); \quad \text{then,}$$
$$\underline{\mathbf{h}}_{\mathbf{sort}} = [0.0854 \quad 0.1256 \quad 0.4133 \quad 0.416 \quad 0.7561$$
$$0.817 \quad 0.9395 \quad 0.9698 \quad 1.0869 \quad 1.1069$$
$$1.5772 \quad 1.6002 \quad 1.6239 \quad 1.8351 \quad 2.496],$$
$$\pi: \quad \underline{\mathbf{Ix}} = [11 \quad 6 \quad 9 \quad 5 \quad 2 \quad 13 \quad 7 \quad 14$$
$$1 \quad 12 \quad 15 \quad 8 \quad 10 \quad 4 \quad 3],$$
$$\underline{\mathbf{r}}_{\mathbf{sort}} = [0 \quad 1 \quad 1 \quad 1 \quad 0 \quad 0 \quad 1 \quad 1 \quad 1 \quad 0$$
$$0 \quad 1 \quad 0 \quad 1 \quad 0].$$

For a BCH(15,7,5) code, the minimum distance is $d_{\min} = 5$. Consequently, $d = d_{\min} - 1 = 4$. Hence the set of flip vectors, denoted as $\Phi$, is thus expressed as follows:

$$\Phi = \{$$
$$[1 \quad 0 \quad 0 \quad 0 \quad 0 \quad 0 \quad 0 \quad 0 \quad 0 \quad 0 \quad 0 \quad 0 \quad 0 \quad 0 \quad 0],$$
$$[0 \quad 1 \quad 0 \quad 0 \quad 0 \quad 0 \quad 0 \quad 0 \quad 0 \quad 0 \quad 0 \quad 0 \quad 0 \quad 0 \quad 0],$$
$$[1 \quad 1 \quad 0 \quad 0 \quad 0 \quad 0 \quad 0 \quad 0 \quad 0 \quad 0 \quad 0 \quad 0 \quad 0 \quad 0 \quad 0],$$
$$[0 \quad 0 \quad 1 \quad 0 \quad 0 \quad 0 \quad 0 \quad 0 \quad 0 \quad 0 \quad 0 \quad 0 \quad 0 \quad 0 \quad 0],$$
$$[1 \quad 0 \quad 1 \quad 0 \quad 0 \quad 0 \quad 0 \quad 0 \quad 0 \quad 0 \quad 0 \quad 0 \quad 0 \quad 0 \quad 0],$$
$$[0 \quad 1 \quad 1 \quad 0 \quad 0 \quad 0 \quad 0 \quad 0 \quad 0 \quad 0 \quad 0 \quad 0 \quad 0 \quad 0 \quad 0],$$
$$[1 \quad 1 \quad 1 \quad 0 \quad 0 \quad 0 \quad 0 \quad 0 \quad 0 \quad 0 \quad 0 \quad 0 \quad 0 \quad 0 \quad 0],$$
$$[0 \quad 0 \quad 0 \quad 1 \quad 0 \quad 0 \quad 0 \quad 0 \quad 0 \quad 0 \quad 0 \quad 0 \quad 0 \quad 0 \quad 0],$$
$$[1 \quad 0 \quad 0 \quad 1 \quad 0 \quad 0 \quad 0 \quad 0 \quad 0 \quad 0 \quad 0 \quad 0 \quad 0 \quad 0 \quad 0],$$
$$[0 \quad 1 \quad 0 \quad 1 \quad 0 \quad 0 \quad 0 \quad 0 \quad 0 \quad 0 \quad 0 \quad 0 \quad 0 \quad 0 \quad 0],$$
$$[1 \quad 1 \quad 0 \quad 1 \quad 0 \quad 0 \quad 0 \quad 0 \quad 0 \quad 0 \quad 0 \quad 0 \quad 0 \quad 0 \quad 0],$$
$$[0 \quad 0 \quad 1 \quad 1 \quad 0 \quad 0 \quad 0 \quad 0 \quad 0 \quad 0 \quad 0 \quad 0 \quad 0 \quad 0 \quad 0],$$
$$[1 \quad 0 \quad 1 \quad 1 \quad 0 \quad 0 \quad 0 \quad 0 \quad 0 \quad 0 \quad 0 \quad 0 \quad 0 \quad 0 \quad 0],$$
$$[0 \quad 1 \quad 1 \quad 1 \quad 0 \quad 0 \quad 0 \quad 0 \quad 0 \quad 0 \quad 0 \quad 0 \quad 0 \quad 0 \quad 0],$$
$$[1 \quad 1 \quad 1 \quad 1 \quad 0 \quad 0 \quad 0 \quad 0 \quad 0 \quad 0 \quad 0 \quad 0 \quad 0 \quad 0 \quad 0],$$
$$\}$$

Let us now compute the syndrome for the received codeword, $\underline{\mathbf{r}}$, resulting in $\underline{\mathbf{S}}_{\mathbf{y}} = [0 \quad 1 \quad 1 \quad 1 \quad 0 \quad 0 \quad 1 \quad 1]$. As the syndrome is non-zero ($\underline{\mathbf{S}}_{\mathbf{y}} \neq \underline{\mathbf{0}}$), we select the first flip vector from the set $\Phi$ and add it to $\underline{\mathbf{r}}_{\mathbf{sort}}$. Subsequently, we calculate the syndrome for the resultant vector as part of the valid codeword check. Thus, we have:

$$\underline{\mathbf{f}}_{\mathbf{pv}} = \Phi(1);$$
$$\underline{\mathbf{r}}_{\mathbf{t}} = \underline{\mathbf{r}}_{\mathbf{sort}} \oplus \underline{\mathbf{f}}_{\mathbf{pv}};$$
$$= [0 \quad 1 \quad 1 \quad 1 \quad 0 \quad 0 \quad 1 \quad 1 \quad 1 \quad 0 \quad 0 \quad 1 \quad 0 \quad 1 \quad 0] \oplus$$
$$[1 \quad 0 \quad 0 \quad 0 \quad 0 \quad 0 \quad 0 \quad 0 \quad 0 \quad 0 \quad 0 \quad 0 \quad 0 \quad 0 \quad 0]$$
$$= [1 \quad 1 \quad 1 \quad 1 \quad 0 \quad 0 \quad 1 \quad 1 \quad 1 \quad 0 \quad 0 \quad 1 \quad 0 \quad 1 \quad 0].$$
$$\underline{\mathbf{r}}_{\mathbf{o}} = \pi^{-1}(\underline{\mathbf{r}}_{\mathbf{t}});$$
$$= [1 \quad 0 \quad 0 \quad 1 \quad 1 \quad 1 \quad 1 \quad 1 \quad 1 \quad 1 \quad 0 \quad 1 \quad 0 \quad 0 \quad 1 \quad 0].$$
$$\underline{\mathbf{S}}_{\mathbf{y}} = \underline{\mathbf{r}}_{\mathbf{o}} \mathbf{H}^{\mathbf{T}};$$
$$= [0 \quad 1 \quad 1 \quad 0 \quad 0 \quad 0 \quad 1 \quad 1].$$

Since the syndrome is non-zero, we proceed by selecting the next flip vector from the set $\Phi$ and repeat the same steps:

$$\underline{\mathbf{f}}_{\mathbf{pv}} = \Phi(2);$$
$$\underline{\mathbf{r}}_{\mathbf{t}} = \underline{\mathbf{r}}_{\mathbf{sort}} \oplus \underline{\mathbf{f}}_{\mathbf{pv}};$$
$$= [0 \quad 1 \quad 1 \quad 1 \quad 0 \quad 0 \quad 1 \quad 1 \quad 1 \quad 0 \quad 0 \quad 1 \quad 0 \quad 1 \quad 0] \oplus$$
$$[0 \quad 1 \quad 0 \quad 0 \quad 0 \quad 0 \quad 0 \quad 0 \quad 0 \quad 0 \quad 0 \quad 0 \quad 0 \quad 0 \quad 0]$$
$$= [0 \quad 0 \quad 1 \quad 1 \quad 0 \quad 0 \quad 1 \quad 1 \quad 1 \quad 0 \quad 0 \quad 1 \quad 0 \quad 1 \quad 0].$$
$$\underline{\mathbf{r}}_{\mathbf{o}} = \pi^{-1}(\underline{\mathbf{r}}_{\mathbf{t}});$$
$$= [1 \quad 0 \quad 0 \quad 1 \quad 1 \quad 0 \quad 1 \quad 1 \quad 1 \quad 0 \quad 0 \quad 0 \quad 0 \quad 1 \quad 0].$$
$$\underline{\mathbf{S}}_{\mathbf{y}} = \underline{\mathbf{r}}_{\mathbf{o}} \mathbf{H}^{\mathbf{T}};$$
$$= [0 \quad 0 \quad 0 \quad 0 \quad 0 \quad 0 \quad 0 \quad 0].$$

With the second iteration yielding a zero syndrome, indicating the validity of $\underline{\mathbf{r}}_{\mathbf{o}}$ as a codeword, the process concludes. The decoded codeword ($\underline{\mathbf{c}}^{\mathbf{opt}}$) is determined as $\underline{\mathbf{r}}_{\mathbf{o}}$, and the associated message bits are extracted by selecting the first $k$ bits from $\underline{\mathbf{c}}^{\mathbf{opt}}$, since we considered the systematic BCH(15,7,5) code.

By systematically exploring and manipulating the least reliable bits, the flip decoder enhances the decoding performance, while maintaining diversity.

The proposed DFD achieves diversity, thereby offering excellent performance over fading channels at a very low complexity. By contrast, the original hard-decision GRAND [53] does not preserve diversity order due to its hard-decision nature, resulting in lower performance compared to DFD. Other soft-decision GRAND variants, such as ORBGRAND [58], exhibit superior performance to DFD at a significantly higher complexity. Additionally, all GRAND variants introduce complexity in two aspects: the complexity of generating the next tentatively assumed noise sequence and the complexity involved in verifying the validity of the resultant codeword through syndrome calculations.

In the case of DFD, the set of flip patterns remains fixed for a given minimum distance ($d_{min}$), regardless of the code. Flip vectors can be generated from this set by appending zeros based on the codeword length ($n$). Consequently, there is no complexity associated with generating a new flip vector each time that needs to be added to the original received word. Thus, the complexity of DFD is significantly lower compared to GRAND and its variants.

To distinguish between the proposed DFD and ORB-GRAND methodologies, the error patterns corresponding to DFD and ORBGRAND in case of (15,7,5) BCH code are illustrated in Figures 1a and 1b, respectively. Figure 1a displays the flip vector pattern associated with DFD, while Figure 1b showcases the error pattern of ORBGRAND up to the 15th query order. In these representations, the black color signifies the flip operation, whereas white indicates the absence of any flip.

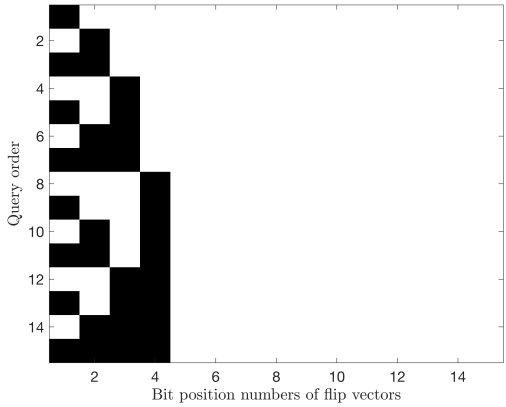

(a) Proposed DFD Flip vector pattern sequence.

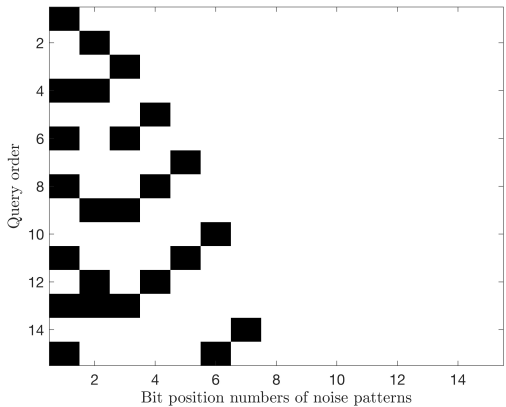

(b) ORBGRAND error pattern sequence

FIGURE 1: Error patterns of DFD and ORBGRAND: BCH (15, 7).

---

**Algorithm 1:** Universal Diversity Flip decoder

**Input:**
- $\Phi$: List of Flip vectors.
- $\underline{\mathbf{r}}$: received codeword.
- $\underline{\mathbf{h}}$: CSI vector.
- $\mathbf{H}$
- $d_{\min}$
- $n_\Phi$: The cardinality of the set $\Phi$.

**Output:**
- $\underline{\mathbf{c}}_o^{\text{opt}}$: The decoded codeword.

1   $\underline{\mathbf{S}}_{\mathbf{y}} = \underline{\mathbf{r}}\mathbf{H}^{\mathbf{T}}$;
2   **if** $(sum(\underline{\mathbf{S}}_{\mathbf{y}}) > 0)$ **then**
3     $d = d_{\min} - 1$;
4     $[\underline{\mathbf{h}}_{\text{sort}}, \pi] = \text{sort}(\underline{\mathbf{h}})$;
5     $\underline{\mathbf{r}}_{\text{sort}} = \pi(\underline{\mathbf{r}})$;
6     $\underline{\mathbf{S}}_{\mathbf{y}} = \underline{\mathbf{r}}\mathbf{H}^{\mathbf{T}}$;
7     $lc = 1$;     // loop counter initialization
8     **while** $(sum(\underline{\mathbf{S}}_{\mathbf{y}}) > 0)$ & $(lc \leq n_\Phi)$ **do**
9       $\underline{\mathbf{f}}_{\mathbf{pv}} = \Phi(l)$;
10      $\underline{\mathbf{r}}_{\mathbf{t}} = \underline{\mathbf{r}}_{\text{sort}} \oplus \underline{\mathbf{f}}_{\mathbf{pv}}$;
11      $\underline{\mathbf{r}}_{\mathbf{o}} = \pi^{-1}(\underline{\mathbf{r}}_{\mathbf{t}})$;
12      $\underline{\mathbf{S}}_{\mathbf{y}} = \underline{\mathbf{r}}_{\mathbf{o}}\mathbf{H}^{\mathbf{T}}$;
13      $lc = lc + 1$;
14     **end**
15   **else**
16     $\underline{\mathbf{r}}_{\mathbf{o}} = \underline{\mathbf{r}}$
17   **end**
18   **if** $(sum(\underline{\mathbf{S}}_{\mathbf{y}}) == 0)$ **then**
19     $\underline{\mathbf{c}}_{\mathbf{o}}^{\text{opt}} = \underline{\mathbf{r}}_{\mathbf{o}}$;
20   **else**
21     $\underline{\mathbf{c}}_{\mathbf{o}}^{\text{opt}} = \underline{\mathbf{r}}$;
22   **end**

---

## IV. The Flip Decoder's Error Probability and Diversity Analysis

We commence by establishing an upper bound for the probability of error denoted as $P_{FD}(\mathscr{C})$. Subsequently, we proceed to deduce the diversity order of the DFD.

Consider the received word $\underline{\mathbf{r}}$ having a total of $n_e$ errors. Additionally, we define $\rho_i = h_i^2 \bar{\rho}_c$ for $1 \leq i \leq n$, representing the instantaneous Signal-to-Noise Ratio (SNR) per coded bit, where $\bar{\rho}_c$ is the average SNR per coded bit.

In the ensuing scenarios, the proposed decoder proves to be unsuccessful in rectifying errors within the received word.

- **Case-1:** In this scenario, where there are $n_e = v$ errors in the received word, with $1 \leq v \leq d$, it is certain that at least one of these errors resides in the Most Reliable Bit (MRB). Consequently, the decoder becomes incapable of rectifying all errors within this context. We denote the probability of this situation by $P_{e_{v}}$.

- **Case-2:** In this particular situation, there are $d + 1$ or more erroneous bits within the received vector $\underline{\mathbf{r}}$. To represent this case, we use $P_{e_{d+1}}$ to denote its probability.

The overall error probability for the proposed DFD is obtained by summing up these individual probabilities. Therefore, we have

$$P_{FD}(\mathscr{C}) = \left(\sum_{v=1}^{d} P_{e_v}\right) + P_{e_{d+1}}. \tag{3}$$

We now proceed with the derivation of the probability of error pertaining to Case-1, wherein the occurrence of $v$ ($1 \leq v \leq d$) errors is considered.

### A. Determination of $\mathbf{P_{e_v}}, 1 \leq \mathbf{v} \leq \mathbf{d}$ :

To compute $P_{e_v}$, we arrange the $\rho_i$ values in ascending order, that is, $\rho_1 \leq \rho_2 \leq \cdots \leq \rho_n$. We introduce the notation $q_i$ as follows:

$$q_i = Prob\{n_i \geq \rho_i | \rho_i\} = Q\left(\sqrt{2\rho_i}\right) \leq e^{-\rho_i},$$

where the final inequality is derived with reference to the Chernoff bound [91]. Let $\bar{q}_i = \mathbf{E}_{\rho_i}[q_i]$. This implies

$$\bar{q}_1 \geq \bar{q}_2 \geq \cdots \geq \bar{q}_n, \tag{4}$$

$$(1 - \bar{q}_1) \leq (1 - \bar{q}_2) \leq \cdots \leq (1 - \bar{q}_n). \tag{5}$$

Initially, we derive the probability of error expressions for both single and double error scenarios. Subsequently, we generalize these expressions to encompass all $v$ errors, where $1 \leq v \leq d$.

### 1) $\mathbf{P_{e_1}}$ derivation

In this scenario, we consider a single error occurring exclusively in the MRB section of $\underline{r}$, i.e. without any errors in the LRB section. The probability of a single error occurring in the $i$-th location is represented by $\bar{q}_i \prod_{j=1}^{n} (1 - \bar{q}_j)$. When we sum over all potential locations within the MRB, we obtain:

$$P_{e_1} = \sum_{i=(d+1)}^{n} \bar{q}_i \prod_{\substack{j=1 \\ j \neq i}}^{n} (1 - \bar{q}_j).$$

Then, $P_{e_1}$ can be upper-bounded as

$$P_{e_1} \leq \sum_{i=(d+1)}^{n} \bar{q}_i (1 - \bar{q}_n)^{n-1}, (using \ (5))$$

$$\leq (n - d)\bar{q}_{d+1} (1 - \bar{q}_n)^{n-1}. (using \ (4)). \tag{6}$$

### 2) $\mathbf{P_{e_2}}$ derivation

This scenario involves the presence of two errors. Among these two errors, at least one resides in the MRB section, which results in decoding failure. Therefore, $P_{e_2}$ comprises two probabilities: one representing the probability of a single error occurring in the MRB ($P_{e_{21}}$), and the other indicating the probability of both errors being in the MRB ($P_{e_{22}}$) yielding:

$$P_{e_2} = P_{e_{21}} + P_{e_{22}}. \tag{7}$$

To calculate $P_{e_{21}}$, we note that the probability of two errors occurring in the $i$-th and $j$-th locations is represented as $\bar{q}_i \bar{q}_j \prod_{\substack{k=1 \\ k \neq i,j}}^{n} (1 - \bar{q}_k)$. By summing over all feasible locations within the MRB section for $j$ and the LRB section for $i$, we obtain:

$$P_{e_{21}} = \sum_{i=1}^{d} \sum_{j=(d+1)}^{n} \bar{q}_i \bar{q}_j \prod_{\substack{k=1 \\ k \neq i,j}}^{n} (1 - \bar{q}_j).$$

We now establish an upper bound for the probability $P_{e_{21}}$ by employing (4) and (5), as follows:

$$P_{e_{21}} \leq \sum_{i=1}^{d} \sum_{j=(d+1)}^{n} \bar{q}_i \bar{q}_j (1 - \bar{q}_n)^{n-2}, (using \ (5))$$

$$= \left(\sum_{i=1}^{d} \bar{q}_i\right) \left[\sum_{j=d+1}^{n} \bar{q}_j\right] (1 - \bar{q}_n)^{n-2},$$

$$\leq (d\bar{q}_1)[(n - d)\bar{q}_{d+1}] (1 - \bar{q}_n)^{n-2}, \quad (using \ (4))$$

$$\leq d(n - d)\bar{q}_1 \bar{q}_{d+1}(1 - \bar{q}_n)^{n-2}. \tag{8}$$

In a similar fashion, we formulate $P_{e_{22}}$ as follows:

$$P_{e_{22}} = Prob\{2 \ errors \ in \ MRB\},$$

$$= \sum_{j=d+1}^{n-1} \sum_{k=j+1}^{n} \bar{q}_j \bar{q}_k \prod_{\substack{l=1 \\ l \neq j,k}}^{n} (1 - \bar{q}_l),$$

$$\leq \sum_{j=d+1}^{n-1} (n - j)\bar{q}_{d+1}^2 (1 - \bar{q}_n)^{n-2} \quad (using \ (4),(5))$$

$$\leq \left(\sum_{j=d+1}^{n-1} n - \sum_{j=d+1}^{n-1} j \bar{q}_{d+1}^2\right) (1 - \bar{q}_n)^{n-2}$$

$$\leq \binom{n-2}{2} \bar{q}_{d+1}^2 (1 - \bar{q}_n)^{n-2}. \tag{9}$$

### 3) $\mathbf{P_{e_v} (1 \leq v \leq d)}$ derivation

Building upon the insights gained from the preceding two subsections, where we calculated the error probabilities for the single and double error cases, we extend our analysis to derive the error probability for scenarios involving $v$ errors. Among these $v$ errors, we denote the number of errors in the MRB section by $v_m$ and the number of errors in the LRB section by $v_n$. Hence, we have:

$$v = v_m + v_n. \tag{10}$$

We denote $P_{e_v}(v_m, v_n)$ as the probability associated with $v_m$ errors in the MRB section and the $v_n$ errors in the LRB section. Consequently, $P_{e_v}$ can be expressed as follows:

$$P_{e_v} = \sum_{v_m=1}^{v} P_{e_v}(v_m, v_n). \tag{11}$$

Let us now derive $P_{e_v}(v_m, v_n)$ as follows:

$$P_{e_v}(v_m, v_n) = \left( \sum_{j_1=d+1}^{n-v_m+1} \sum_{j_2=j_1+1}^{n-v_m} \cdots \sum_{j_{v_m}=j_{(v_{m}-1)}+1}^{n} \bar{q}_{j_1} \bar{q}_{j_2} \cdots \bar{q}_{j_{v_m}} \right)$$

$$\left( \sum_{k_1=1}^{d} \sum_{k_2=k_1+1}^{d-1} \cdots \sum_{k_{v_n}=k_{(v_{n}-1)}+1}^{d-v_n+1} \bar{q}_{k_1} \bar{q}_{k_2} \cdots \bar{q}_{k_{v_n}} \right)$$

$$\left( \prod_{\substack{v=1 \\ v \neq j_1, j_2, \cdots j_{v_m} \\ v \neq k_1, k_2, \cdots k_{v_n}}}^{n} (1 - \bar{q}_v) \right),$$

$$\leq \left( \binom{n-d}{v_m} \bar{q}_{d+1} \bar{q}_{d+2} \cdots \bar{q}_{d+v_m} \right)$$

$$\left( \binom{d}{v_n} \bar{q}_1 \bar{q}_2 \cdots \bar{q}_{v_n} \right) \left( (1 - \bar{q}_n)^{n-v} \right), \text{(using (5))}$$

$$\leq \left( \binom{n-d}{v_m} (\bar{q}_{d+1})^{v_m} \right)$$

$$\left( \binom{d}{v_n} (\bar{q}_1) \right)^{v_n} \left( (1 - \bar{q}_n)^{n-v} \right), \text{(using (4))}$$

$$\leq \binom{n-d}{v_m} \binom{d}{v_n} (\bar{q}_{d+1})^{v_m} (\bar{q}_1)^{v_n} (1 - \bar{q}_n)^{n-v}. \quad (12)$$

Upon substituting (12) into (11), we arrive at:

$$P_{e_v} \leq \sum_{v_m=1}^{v} \binom{n-d}{v_m} \binom{d}{v_n} (\bar{q}_{d+1})^{v_m} (\bar{q}_1)^{v_n} (1 - \bar{q}_n)^{n-v}. \quad (13)$$

#### 4) Derivation of $\bar{q}_1$, $\bar{q}_{d+1}$ and $\bar{q}_n$

Observe that the upper bound in (13) relies on the values of $\bar{q}_1$, $\bar{q}_{d+1}$, and $\bar{q}_n$. To derive expressions for these values, we must delve into the associated ordered statistics.

Consider a continuous random variable $Y$ having a probability density function (PDF) $f(Y)$ and cumulative distribution function (CDF) $F(Y)$. Let $(Y_1, Y_2, \cdots, Y_n)$ represent a random sample, and $(Y_{(1)}, Y_{(2)}, \cdots, Y_{(n)})$ denote the corresponding sample sorted in ascending order so that $Y_{(1)} < Y_{(2)} < \cdots < Y_{(n)}$. The PDF of the $r^{th}$ order statistics, $Y_{(r)}$ [92], can be expressed as:

$$f_{Y_{(r)}} = \frac{n!}{(r-1)!(n-r)!} [F(y)]^{r-1} [1-F(y)]^{n-r} f(y). \quad (14)$$

Let us now define $a = \frac{n!}{(r-1)!(n-r)!}$, and let $Y_i$ be represented by $\rho_i$, where a gain $\rho_i = h_i^2 \bar{\rho}_c$ denotes the instantaneous SNR and follows an exponential distribution. Upon substituting the corresponding CDF $F(Y)$ and PDF

$f(y)$ into (14), we obtain:

$$f_{Y_{(r)}} = a \left( 1 - e^{-\frac{y}{\bar{\rho}_c}} \right)^{r-1} \left( e^{-\frac{y}{\bar{\rho}_c}} \right)^{n-r} \left( \frac{1}{\bar{\rho}_c} e^{-\frac{y}{\bar{\rho}_c}} \right),$$

$$= \frac{a}{\bar{\rho}_c} \left( e^{-\frac{y}{\bar{\rho}_c}} \right)^{n-r+1} \left( 1 - e^{-\frac{y}{\bar{\rho}_c}} \right)^{r-1},$$

$$= \frac{a}{\bar{\rho}_c} \sum_{m=0}^{r-1} \binom{r-1}{m} (-1)^m e^{-\frac{my}{\bar{\rho}_c}} e^{-\frac{y(n-r+1)}{\bar{\rho}_c}},$$

$$= \frac{a}{\bar{\rho}_c} \sum_{m=0}^{r-1} \binom{r-1}{m} (-1)^m e^{-\frac{y}{\bar{\rho}_c}(n-r+1+m)}. \quad (15)$$

To determine $\bar{q}_r$, we calculate its expectation using (15). This leads us to the following expression:

$$\bar{q}_r = \int_0^\infty \frac{a}{\bar{\rho}_c} \sum_{m=0}^{r-1} \binom{r-1}{m} (-1)^m e^{-\frac{y}{\bar{\rho}_c}(n-r+1+m)} e^{-y} dy,$$

$$= \int_0^\infty \frac{a}{\bar{\rho}_c} \sum_{m=0}^{r-1} \binom{r-1}{m} (-1)^m e^{-\frac{y}{\bar{\rho}_c}(n-r+1+m+\bar{\rho}_c)},$$

$$= a \left( \sum_{m=0}^{r-1} \binom{r-1}{m} \frac{(-1)^m}{n-r+1+m+\bar{\rho}_c} \right),$$

$$= \frac{a(r-1)!}{\prod_{k=0}^{r-1} (\bar{\rho}_c + n - k)},$$

$$= \frac{(n!/(n-r)!)}{\prod_{k=0}^{r-1} (\bar{\rho}_c + n - k)},$$

$$= \frac{\prod_{j=0}^{r-1} (n-j)}{\prod_{k=0}^{r-1} (\bar{\rho}_c + n - k)},$$

$$= \frac{^nP_r}{\prod_{k=0}^{r-1} (\bar{\rho}_c + n - k)}. \quad (16)$$

By substituting the values $r = 1, r = (d+1)$, and $r = n$ into (16), we arrive at

$$\bar{q}_1 = \frac{n}{n + \bar{\rho}_c}. \quad (17)$$

$$\bar{q}_{d+1} = \frac{^nP_{d+1}}{\prod_{k=0}^{d} (\bar{\rho}_c + n - k)}. \quad (18)$$

$$\bar{q}_n = \frac{1}{\prod_{k=0}^{n-1} (\bar{\rho}_c + n - k)}. \quad (19)$$

#### 5) Derivation of $P_{e_v}(1 \leq v \leq d)$

By substituting equations (17), (18), and (19) into (13), we arrive at:

$$P_{e_v} \leq \sum_{v_m=1}^{l} \binom{n-d}{v_m} \binom{d}{v_n} \left( \frac{^nP_{d+1}}{\prod_{k=0}^{d} (\bar{\rho}_c + n - k)} \right)^{v_m}$$

$$\left( \frac{n}{n+\bar{\rho}_c} \right)^{v_n} \left( 1 - \frac{1}{\prod_{k=0}^{n-1} (\bar{\rho}_c + n - k)} \right)^{n-v}. \quad (20)$$

### B. Derivation of $P_{e_{(d+1)}}$:

This scenario arises when the number of errors in the received word exceeds the limit of $d = (d_{\min} - 1)$. The proposed decoder is capable of correcting up to $d$ errors since it allows a maximum of $d = (d_{\min} - 1)$ flips in the decoding process. In such cases, the $\rho_i$ values are unordered, and the probability of encountering $d_{\min}$ or more errors can be formulated as

$$P_{e_{(d+1)}} = \sum_{m=d_{\min}}^{n} \binom{n}{m} \bar{q}^m (1 - \bar{q})^{n-m}. \qquad (21)$$

Here, $\bar{q}$ represents the average probability of bit error associated with the unordered $\rho_i$ values. In terms of the average SNR, $\bar{q}$ can be expressed as:

$$\bar{q} = \mathbf{E}_\rho \left[ Prob\{n \geq \rho | \rho\} \right] = \mathbf{E}_\rho \left[ Q\left(\sqrt{2\rho}\right) \right] \leq \frac{1}{1 + \bar{\rho}_c}. \qquad (22)$$

We now express the upper bound on $\bar{q}$ by applying the Chernoff bound [91], followed by averaging over the distribution of $\rho$ upon substituting (22) into (21), we arrive at:

$$P_{e_{(d+1)}} = \sum_{m=(d+1)}^{n} \binom{n}{m} \left\{ \frac{1}{1 + \bar{\rho}_c} \right\}^m \left\{ \frac{\bar{\rho}_c}{1 + \bar{\rho}_c} \right\}^{n-m}. \qquad (23)$$

We are now ready to establish an upper bound on the total average probability of error by incorporating (20) and (23) into (3), which yields (24). The upper bound shown by the equation (24) is lose bound but gives the same diversity order.

### C. The Diversity Flip Decoder's Diversity order
Let us now proceed by determining the diversity order of the DFD.

### Theorem 1:
The DFD achieves a diversity order of $d_{\min}$ for a binary linear block code $\mathscr{C}(n, k)$ having a minimum Hamming distance of $d_{\min}$.

### Proof:
Referring to an equation (24), let

$$\mathscr{A} = \sum_{v=1}^{d} \sum_{v_m=1}^{v} \binom{n-d}{v_m} \binom{d}{v_n} \left( \frac{^n P_{d+1}}{\prod_{k=0}^{d}(\bar{\rho}_c + n - k)} \right)^{v_m}$$
$$\left( \frac{n}{n + \bar{\rho}_c} \right)^{v_n} \left( 1 - \frac{1}{\prod_{k=0}^{n-1}(\bar{\rho}_c + n - k)} \right)^{n-v} \qquad (25)$$

$$\mathscr{B} = \sum_{m=(d+1)}^{n} \binom{n}{m} \left\{ \frac{1}{1 + \bar{\rho}_c} \right\}^m \left\{ \frac{\bar{\rho}_c}{1 + \bar{\rho}_c} \right\}^{n-m}$$

$$\mathscr{B} = \left[ \binom{n}{d+1} \left\{ \frac{1}{1 + \bar{\rho}_c} \right\}^{(d+1)} \left\{ \frac{\bar{\rho}_c}{1 + \bar{\rho}_c} \right\}^{n-(d+1)} \right] +$$
$$\left[ \binom{n}{d+2} \left\{ \frac{1}{1 + \bar{\rho}_c} \right\}^{(d+2)} \left\{ \frac{\bar{\rho}_c}{1 + \bar{\rho}_c} \right\}^{n-(d+2)} \right] + \cdots$$
$$\cdots \cdots + \left[ \binom{n}{n} \left\{ \frac{1}{1 + \bar{\rho}_c} \right\}^n \right]. \qquad (26)$$

Then, Equation (24) can be written as

$$P_{FD}(\mathscr{C}) = \mathscr{A} + \mathscr{B}. \qquad (27)$$

By invoking the definition of diversity order as outlined in [23] in the limit (27) can be expressed as

$$D = \lim_{\bar{\rho}_c \to \infty} \frac{-\log[P_{FD}(\mathscr{C})]}{\log[\bar{\rho}_c]}$$
$$= \lim_{\bar{\rho}_c \to \infty} \frac{-\log[\mathscr{A} + \mathscr{B}]}{\log[\bar{\rho}_c]}.$$

But, as $\bar{\rho}_c \to \infty$, the term $\mathscr{A} \to 0$, hence:

$$D = \lim_{\bar{\rho}_c \to \infty} \frac{-\log[\mathscr{B}]}{\log[\bar{\rho}_c]}.$$

As $\bar{\rho}_c \to \infty$, the exponent $(d+1)$ starts to dominate over other exponentials in the term B (equation (26)). Therefore we have

$$D = d + 1$$
$$= d_{\min}. \qquad (28)$$

∎

### D. Computational Complexity Analysis
The DFD combines the benefits of low complexity with the preservation of diversity. It relies on a comprehensive list of flip vectors denoted as $\underline{\mathbf{f}}_{\mathbf{pv}}$, which encompasses a total of $2^d - 1$ ($2^{d_{\min} - 1} - 1$) vectors. The computational complexity of the DFD is directly influenced by the minimum distance ($d_{\min}$) of the code $\mathscr{C}(n, k)$. Through a systematic search process within the list $\Phi$, the algorithm aims for identifying a suitable flip vector by sequentially evaluating each vector. Even in the worst-case scenario, where Algorithm 1 examines the last flip vector in $\Phi$, the DFD maintains its efficiency, given its worst-case complexity order of $2^{d_{\min} - 1} - 1$. Therefore, this technique has the twin benefits of low computational complexity, while preserving diversity.

Table III provides a detailed comparison of the maximum complexities associated with various decoders.

### V. Simulation Results - DFD
To evaluate the performance of the proposed DFD, several examples are considered. The simulations are conducted in an uncorrelated Rayleigh fading channel with unit variance, using BPSK modulation throughout.

To illustrate the effectiveness of the DFD technique, Bit Error Rate (BER) vs signal-to-noise ratio per message bit ($E_b/N_0$) plots are generated. It should be noted that $\bar{\rho}_c$ and

$$P_{FD}(\mathscr{C}) \leq \sum_{v=1}^{d} \sum_{v_m=1}^{v} \binom{n-d}{v_m} \binom{d}{v_n} \left( \frac{{}^n P_{d+1}}{\prod_{k=0}^{d}(\bar{\rho}_c + n - k)} \right)^{v_m} \left( \frac{n}{n + \bar{\rho}_c} \right)^{v_n} \left( 1 - \frac{1}{\prod_{k=0}^{n-1}(\bar{\rho}_c + n - k)} \right)^{n-v}$$
$$+ \sum_{m=(d+1)}^{n} \binom{n}{m} \left\{ \frac{1}{1 + \bar{\rho}_c} \right\}^m \left\{ \frac{\bar{\rho}_c}{1 + \bar{\rho}_c} \right\}^{n-m} \tag{24}$$

TABLE III: Decoder Complexity: A Comparative Study

| Decoder | Maximum complexity |
|---|---|
| Chase-1 | $\binom{n}{\lfloor \frac{d_{min}}{2} \rfloor} \times$ binary decoder complexity |
| Chase-2 | $2^{\lfloor \frac{d_{min}}{2} \rfloor} \times$ binary decoder complexity |
| Chase-3 | $\lfloor (\frac{d_{min}}{2}) + 1 \rfloor \times$ binary decoder complexity |
| OSD | $\mathcal{O}(k^l)$ |
| ISD | $\mathcal{O}(k^l)$ |
| GRAND | $b$ (Threshold) |
| ORBGRAND | $b$ (Threshold) |
| Proposed DFD | $2^{d_{min}-1} - 1$ |

$E_b/N_0$ are related by $E_b/N_0 = \bar{\rho}_c \times code\ rate$. To achieve a desired BER on the order of $10^{-a}(a \in \mathbb{Z}^+)$, a transmission of $10^{(a+2)}$ bits is performed for the BER simulation plots.

The subsequent section showcases the versatile capability of the proposed Diversity Flip decoder (DFD) in achieving diversity at a low complexity for various code types, thereby establishing its universality. Additionally, we conduct a thorough comparison of the diversity flip decoder's performance vs complexity against those of existing decoding methods in the following subsection.

### A. Simulations: DFD Preserves Diversity

In order to illustrate the performance and diversity preservation capability of the proposed DFD, the simulation results are presented for four specific codes: Parity check codes, Hamming codes, BCH codes, and Polar codes. However, it is important to emphasize that the DFD applies to all types of linear codes.

- **Example-1:-Parity check codes:** Our initial evaluation is focused on Parity check (PC) codes, which exhibit a minimum distance of 2. When a parity mismatch arises at the receiver, the DFD technique acts by toggling the LRB bit associated with the lowest Channel State Information (CSI). Table IV provides the detailed parameters for the PC codes employed in the simulations.
Figure 2 provides the BER curves of the PC $(4, 3)$ and PC $(12, 11)$ as representative examples, which benchmarked both against HDD (hard-decision decoding) and SDD (soft-decision decoding). At high SNR levels, the DFD curve mirrors the slope of the SDD curve, indicating an equivalent level of diversity capability to that of soft-decision decoding. This observation clearly demonstrates that, as shown in Figure 2, the DFD technique achieves a diversity order of 2 for PC codes operating in a Rayleigh fading channel. Furthermore, Figure 3 extends our performance evaluations of DFD in comparison to hard-decision decoding for the PC $(64, 63)$, $(128, 127)$, and $(256, 255)$ codes. These results indicate that DFD attains a second-order diversity for PC codes.
It is worth noting that PC codes possess a minimum distance of $d_{min} = 2$, thus lacking error-correcting capabilities (t = 0). As a result, PC codes are primarily employed for error detection purposes. However, the proposed DFD technique exhibits the capability of correcting single errors with high probability in the presence of fading, despite having $d_{min} = 2$.

- **Example-2:- Hamming codes:** Next, we examine the performance of the DFD technique for Hamming codes, which offer a minimum distance of 3. In this case, two bits associated with the two lowest CSI values of the received word are selected for flipping, as per Algorithm 1. There are three possible ways of flipping these two bits, until a valid codeword is obtained. The parameters of the Hamming codes used for the simulations are provided in Table V.
Figure 4 presents a comparison of the DFD to HDD and SDD for the $(7, 4)$ and $(15, 11)$ Hamming codes. At high Signal-to-Noise Ratio (SNR), both the DFD curve and the SDD curve exhibit the same slope for both the $(7, 4)$ and $(15, 11)$ Hamming codes, indicating a diversity order of 3, which is equivalent to that of soft-decision decoding. Additionally, the figure displays the curves for hard-decision decoding and the DFD used for the $(127, 120)$ and $(255, 247)$ Hamming codes, further confirming that the DFD technique achieves a diversity order of 3 for Hamming codes.

- **Example-3:- BCH codes:** We now turn our attention to BCH codes. The simulation parameters for the corresponding BCH codes are presented in Table VI.
Figure 5 presents the performance of the DFD technique applied to BCH codes. The plot offers a comparative analysis involving the DFD technique, HDD, and SDD for the $(15, 7)$ BCH code. At high SNR values, the DFD mirrors the slope of the SDD curve, demonstrating a diversity order equivalent to $d_{min}$ (minimum distance). Furthermore, the figure extends the evaluation by including the curves of the DFD and of the hard decision aided $(255, 239)$, $(63, 51)$, and $(31, 21)$ BCH codes, conclusively confirming that the proposed DFD attains a diversity order of $d_{min}$.
It is worth highlighting that the IEEE 802.15.6 standard of Wireless Body Area Networks (WBAN) [93] has chosen to employ short BCH codes, specifically those with block lengths of $n = 63$ and $n = 31$. The proposed DFD outperforms the classic Peterson algorithm in the Rayleigh fading channel while preserving diversity. Moreover, it is important to emphasize that the DFD technique demonstrates reduced computational complexity in comparison to the Peterson algorithm, a point that will be elaborated on in the following discussion. Peterson's algorithm and the Berlekamp-Massey algorithm (BM) are commonly utilized for the decoding of BCH codes. Peterson's algorithm exhibits a worst-case complexity of $2t(n-1)^2 + \Theta(t^3)$, whereas the Berlekamp-Massey algorithm (BM) has a worst-case complexity of $2t(n-1)^2 + \Theta(t^2)$ [94]. By contrast, the worst-case complexity of the proposed DFD, expressed as $2^{(d_{min}-1)} - 1$, is substantially lower.

TABLE IV: Parameters of PC Codes

| (n, k) | $d_{min}$ | Correcting capability, t | Maximum No. of Correctable Errors with FD |
|--------|-----------|--------------------------|-------------------------------------------|
| (4,3) | 2 | 0 | 1 |
| (12,11) | 2 | 0 | 1 |
| (64,63) | 2 | 0 | 1 |
| (128,127) | 2 | 0 | 1 |
| (256,255) | 2 | 0 | 1 |

TABLE V: Parameters of Hamming Codes

| (n, k) | $d_{min}$ | Correcting capability, t | Maximum No. of Correctable Errors with FD |
|--------|-----------|--------------------------|-------------------------------------------|
| (7,4) | 3 | 1 | 2 |
| (15,11) | 3 | 1 | 2 |
| (127,120) | 3 | 1 | 2 |
| (255,247) | 3 | 1 | 2 |

TABLE VI: Parameters of BCH codes

| (n, k) | $d_{min}$ | Correcting capability, t | Maximum No. of Correctable Errors with FD |
|--------|-----------|--------------------------|-------------------------------------------|
| (15,7) | 5 | 2 | 4 |
| (31,21) | 5 | 2 | 4 |
| (63,51) | 5 | 2 | 4 |
| (255,239) | 5 | 2 | 4 |

TABLE VII: Parameters of Polar Codes

| (n, k) | $d_{min}$ | Maximum No. of Correctable Errors with FD |
|--------|-----------|-------------------------------------------|
| (128,120) | 2 | 1 |
| (128,113) | 4 | 3 |
| (128,106) | 4 | 3 |
| (128,99) | 4 | 3 |

TABLE VIII: Coding gain offered by DFD in case of Hamming codes over uncorrelated Rayleigh fading channel

| $n$ | $k$ | Code Rate (R) | $E_b/N_0$ at BER of | | Coding Gain at BER of | |
|-----|-----|------|-----------|-----------|-----------|-----------|
| | | | $10^{-5}$ | $10^{-6}$ | $10^{-5}$ | $10^{-6}$ |
| 7 | 4 | 0.57 | 19dB | 22.2dB | 24.9dB | 31.8dB |
| 15 | 11 | 0.73 | 20.2dB | 23.6dB | 23.7dB | 30.4dB |
| 31 | 26 | 0.83 | 21.7dB | 25.2dB | 22.2dB | 28.8dB |
| 63 | 57 | 0.9 | 23.4dB | 26.8dB | 20.5dB | 27.2dB |
| 127 | 120 | 0.94 | 25dB | 28.6dB | 18.9dB | 25.4dB |

- **Example-4:- Polar codes:** Finally, the family of Polar codes was examined to evaluate the performance of the DFD. Various Polar codes were selected for simulation, and the simulation parameters are provided in Table VII. The minimum distance ($d_{min}$) of the Polar codes was determined using the $gfweight()$ function in MATLAB. The BER performance using the proposed DFD is illustrated in Figure 6. Interestingly, the Polar codes $(128,113)$, $(128,106)$, and $(128,99)$ exhibit the same minimum distance of $d_{min} = 4$. Furthermore, the BER curves for these codes exhibit identical slopes, indicating the same diversity of $d_{min} = 4$.

It is worth noting that the performance of the original GRAND technique [53] aligns with that of HDD. By contrast, the proposed DFD outperforms the original GRAND at high SNRs, while maintaining exceptionally low computational complexity, as seen later in Figure 9.

### B. Simulations: Performance analysis of DFD

To assess the performance of the proposed DFD, the coding gains attained at BER of $10^{-6}$ are tabulated in Tables VIII and IX for different code lengths and code rates in case of Hamming and BCH codes, respectively. Figure 7 shows the plot of coding gain versus code rate in the case of both BCH and Hamming codes for different code lengths and code rates.

### C. Simulation-Based Comparative Analysis: Performance versus Complexity of Proposed DFD versus Existing Techniques

In order to assess the performance versus complexity of the proposed diversity flip decoder in comparison to existing techniques, the BCH (127,113) code was selected as a case study. Figure 8 showcases the BER performance comparison between the DFD decoder and several existing approaches, such as hard-decision GRAND [54], soft-decision ORBGRAND [57], ORBGRAND with pseudo-soft-information [61], and fading-GRAND [60]. In all the simulations conducted involving GRAND and its variants, the Abandonment Bit value (ABB) is set to $10^6$. In the context of fading-GRAND, the threshold $\Delta$ is determined by assigning $m = -0.02165$ and $b = 0.7924$, as stated in [60]. Observe in Figure 8 that for $E_b/N_o > 20$ dB the proposed DFD performs better than the original hard-decision GRAND.

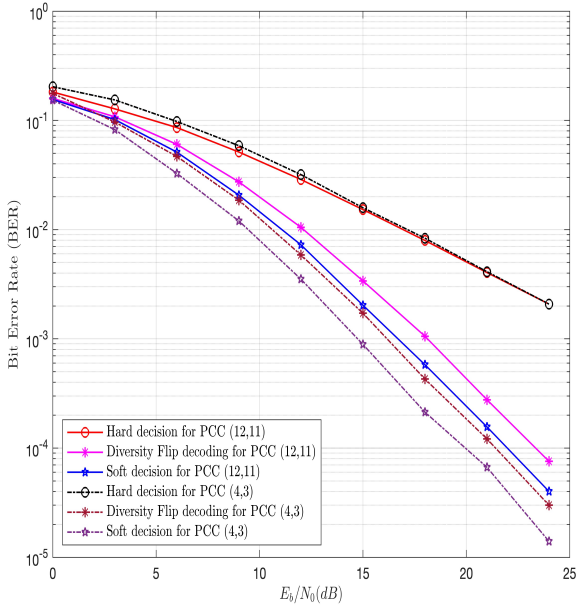

**FIGURE 2:** BER of various short PC codes using the proposed DFD.

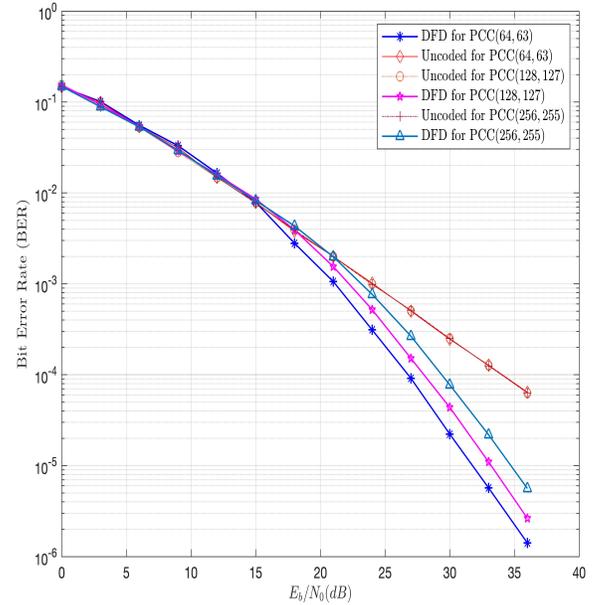

**FIGURE 3:** BER of various long PC codes using the proposed DFD.

The GRAND and its variants such as ORBGRAND [58], fading-GRAND [60], and ORBGRAND [61] associated with pseudo-soft-information, have the following pair of dominant complexity contributions. Firstly, the generation of subsequent tentative noise sequences contributes to the overall complexity. Secondly, the number of codebook accesses required for validating the codeword obtained or calculating the syndrome equation adds to the complexity. By contrast, DFD does not have to generate the next flip vector, since the set of flip vectors is already available at the receiver for a given minimum distance $d_{min}$ and codeword length $n$. The DFD only involves syndrome calculations or codebook queries, hence imposing a low complexity. Therefore, to facilitate a comparison between the DFD and other methods, the number of codebook queries is taken into consideration for complexity comparison throughout this paper. However, the actual overall complexity of GRAND and its variants exceeds the complexity shown in the results of this paper.

The complexity of the decoding algorithms varies depending on both the number and location of errors in the received codeword. Consequently, the overall complexity of decoders exhibits variations versus the SNR values. Since the worst-case complexity determines the maximum hardware-dimension and power-consumption in system design, we present the worst-case complexities in Figure 9. The results clearly demonstrate that the proposed DFD exhibits significantly lower worst-case complexities than the alternative methods.

Figure 10 illustrates the coding gain at a bit error rate (BER) of $10^{-6}$ against worst-case complexity for the DFD, GRAND and its variants. Observe that the DFD achieves an approximately 2 dB higher coding gain than the original hard-decision GRAND at 4 orders lower complexity.

To underscore the minimal complexity of the DFD in comparison to alternative counterparts, Table X tabulates the percentage complexities of the various decoders. The complexities are computed relative to that of the GRAND [53] at BER of $10^{-5}$. Regardless, of whether we consider the maximum (worst-case) complexity or average complexity, the DFD exhibits significantly lower complexity than the other decoders.

The GRAND decoder, known for its hard-decision decoding approach, is outperformed by the proposed DFD, which exhibits superior performance at a modest computational complexity. Although ORBGRAND (soft-decision decoder), fading-GRAND, and ORBGRAND relying on pseudo-soft information provide improved performance compared to the DFD technique, the DFD has the edge in terms of low complexity. Hence the DFD is an attractive choice for light-weight real-world applications.

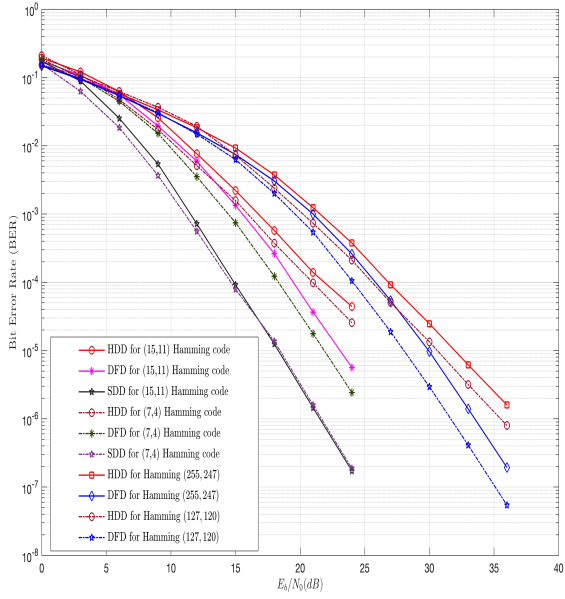

FIGURE 4: BER of various Hamming codes using the proposed DFD.

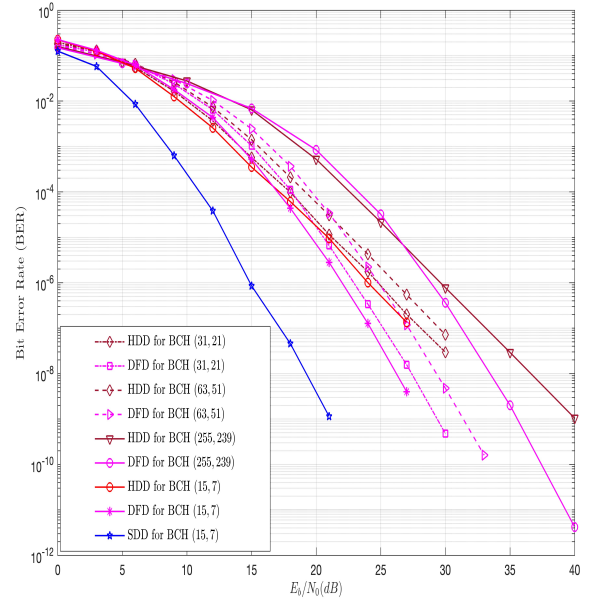

FIGURE 5: BER of various BCH codes using the proposed DFD.

## VI. Extended Diversity Flip Decoder

The Diversity Flip decoder, proposed in Section III flips minimally $d_{min} - 1$ least reliable bits in diverse configurations, thereby maintaining the diversity order across fading channels. In this section, we introduce an Extended Diversity Flip decoder (EDFD), which extends the window of the least reliable bits to be flipped. Moreover, in EDFD, the flip vector corresponding to a valid codeword having the minimum Hamming weight is chosen as the final error vector for obtaining the decoded codeword.

Let $\epsilon$ be the extension window length of the EDFD and denote the flip vector by $\underline{\mathbf{fe}}_{\mathbf{pv}}^{\mathbf{n}}$. Further more, let the set of flip vectors be represented by $\Phi_e$, which is defined as follows:

$$\Phi_e = \left\{ \underline{\mathbf{fe}}_{\mathbf{pv}}^{\mathbf{n}} \ \middle| \ \underline{\mathbf{fe}}_{\mathbf{pv}}^{\mathbf{n}} = \begin{bmatrix} \underline{\mathbf{fep}}_{\mathbf{i}}^{\mathbf{d}+\epsilon} & \mathbf{0}^{\mathbf{n-d-\epsilon}} \end{bmatrix} \right\}$$
$$(1 \leq i \leq d).$$

In this context, $\underline{\mathbf{fep}}_{\mathbf{i}}^{\mathbf{d}+\epsilon}$ represents the flip pattern of the EDFD, characterized by a length of $d + \epsilon$ and a Hamming weight of $i$. For instance, for a $(7, 4, 3)$ Hamming code, we have $d_{min} = 3$, thus $d = d_{min} - 1 = 2$. Subsequently, the flip vector set, denoted as $\Phi_e$, associated with $\epsilon = 1$ is elucidated below:

$$\Phi = \{$$
$$[1 \quad 0 \quad 0 \quad 0 \quad 0 \quad 0 \quad 0],$$
$$[0 \quad 1 \quad 0 \quad 0 \quad 0 \quad 0 \quad 0],$$
$$[0 \quad 0 \quad 1 \quad 0 \quad 0 \quad 0 \quad 0],$$
$$[1 \quad 1 \quad 0 \quad 0 \quad 0 \quad 0 \quad 0],$$
$$[1 \quad 0 \quad 1 \quad 0 \quad 0 \quad 0 \quad 0],$$
$$[0 \quad 1 \quad 1 \quad 0 \quad 0 \quad 0 \quad 0],$$
$$\}$$

The difference between DFD and EDFD is in their respective flip vector lists. The decoding procedure of EDFD mirrors that of DFD, as detailed below. Initially, the Channel State Information (CSI), denoted as $\underline{\mathbf{h}}$ is subjected to sorting to yield $\underline{\mathbf{h}}_{\mathbf{sort}}$, alongside its corresponding permutation function, $\pi_1$. Subsequently, $\pi_1$ is applied to the received codeword $\underline{\mathbf{r}}$ to obtain $\underline{\mathbf{r}}_{\mathbf{sort}}$. The first $(d+\epsilon)$ bits within $\underline{\mathbf{r}}_{\mathbf{sort}}$ represent the Least Reliable Bits (LRB), while the remaining bits constitute the Most Reliable Bits (MRB). Following this, the initial flip vector from the set $\Psi_e$ is selected and added to $\underline{\mathbf{r}}_{\mathbf{sort}}$, yielding $\underline{\mathbf{r}}_{\mathbf{t}}$. The inverse of $\pi_1$, denoted as $\pi_1^{-1}$, is then applied to $\underline{\mathbf{r}}_{\mathbf{t}}$ to recover $\underline{\mathbf{r}}_{\mathbf{o}}$. Subsequently, the syndrome for $\underline{\mathbf{r}}_{\mathbf{o}}$ is computed to ascertain the codeword's validity. If a non-zero syndrome is obtained, the process iterates by selecting the subsequent flip vector from the $\Psi_e$ list. This iteration is continued until a zero syndrome is arrived at. Upon obtaining a zero syndrome, the process terminates, with $\underline{r}_o$ representing the decoded codeword. The EDFD technique is formally summarized in Algorithm 2.

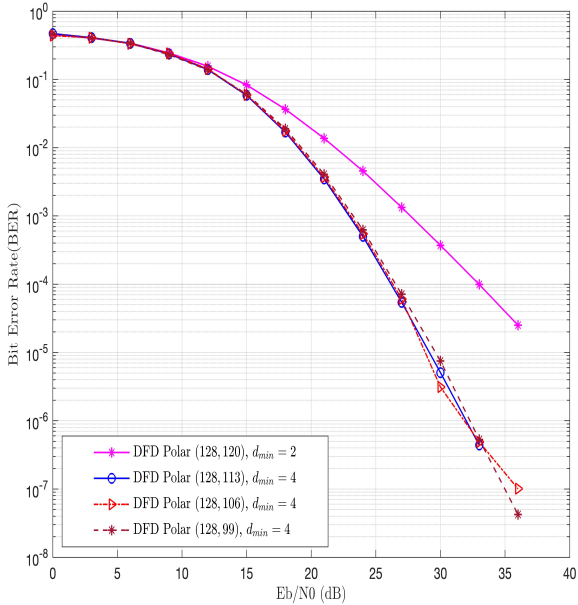

FIGURE 6: BER of various Polar codes using the proposed DFD.

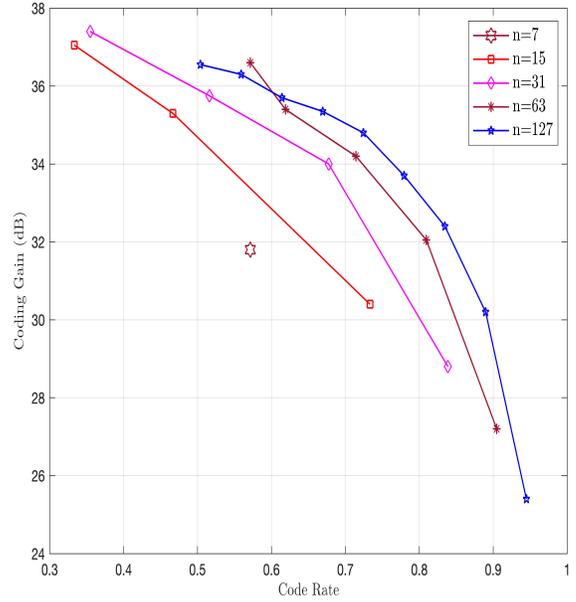

FIGURE 7: Coding Gain of DFD at BER of $10^{-6}$ for different code lengths and code rates in case of Hamming and BCH codes in uncorrelated Rayleigh fading channel.

The flip vector patterns of EDFD for the (15,7,5) BCH code when $\epsilon = 0, 1, 2, 3$ are depicted in Figures 11, 12, 13, and 14 correspondingly.

## VII. Simulation Results: EDFD

This section presents the simulation results for the Extended Diversity Flip Decoder (EDFD) conducted under the same experimental setup as that of the Diversity Flip Decoder (DFD). The simulations are conducted in a Rayleigh fading channel having unit variance, employing Binary Phase Shift Keying (BPSK) modulation. Figure 15 illustrates the Bit Error Rate (BER) simulation outcomes of the EDFD applied to a BCH $(15, 7)$ code for varying values of the extension window length $\epsilon$. It is observed that as the value of $\epsilon$ increases, there is an enhancement in the BER performance. The coding gains offered by EDFD for different values of $\epsilon$ in the case of BCH(15,7) are tabulated in Table XI. Figure 16 displays a scatter plot illustrating the coding gain of EDFD for various values of $\epsilon$ at $E_b/N_0$ of $10^{-5}$. Figure 17 illustrates the BER performance against the worst-case complexity of EDFD for various $\epsilon$ values. EDFD provides varying BER performance levels for different complexities, depending on the chosen $\epsilon$.

## VIII. Conclusions and Discussion

The coding gains at BERs of $10^{-5}$ and $10^{-6}$, as detailed in Table XII, highlight the performance of the proposed DFD

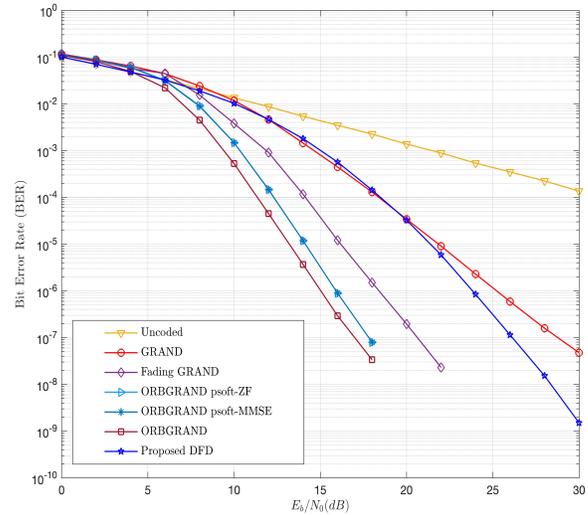

FIGURE 8: BER Comparison of the proposed DFD to existing decoders: $BCH(127, 113)$.

and other decoders for the BCH $(127, 113)$ code. Specifically, the proposed DFD exhibits a coding gain advantage of $0.6\text{dB}$ and $1.4\text{dB}$ over GRAND [53] at BERs of $10^{-5}$ and $10^{-6}$, respectively, despite its significantly lower complexity.

In pursuit of improved BER performance, we introduce the Extended Diversity Flip Decoder (EDFD). Table XI de-

TABLE IX: Coding gain offered by DFD in case of BCH codes over uncorrelated Rayleigh fading channel

| $n$ | $k$ | Code Rate (R) | $E_b/N_0$ at BER of | | Coding Gain at BER of | |
|---|---|---|---|---|---|---|
| | | | $10^{-5}$ | $10^{-6}$ | $10^{-5}$ | $10^{-6}$ |
| 15 | 7 | 0.46 | 16.1dB | 18.7dB | 27.8dB | 35.3dB |
| | 5 | 0.33 | 15.2dB | 16.95dB | 28.7dB | 37.05dB |
| 31 | 21 | 0.67 | 17.8dB | 20dB | 26.1dB | 34dB |
| | 16 | 0.51 | 16.4dB | 18.25dB | 27.5dB | 35.75dB |
| | 11 | 0.35 | 14.9dB | 16.6dB | 29dB | 37.4dB |
| 63 | 51 | 0.83 | 19.3dB | 21.95dB | 24.6dB | 32.05dB |
| | 45 | 0.9 | 17.6dB | 19.8dB | 26.3dB | 34.2dB |
| | 39 | 0.9 | 16.7dB | 18.6dB | 27.2dB | 35.4dB |
| | 36 | 0.9 | 15.7dB | 17.4dB | 28.2dB | 36.6dB |
| 127 | 113 | 0.89 | 21.4dB | 23.8dB | 22.5dB | 30.2dB |
| | 106 | 0.83 | 19.4dB | 21.6dB | 24.5dB | 32.4dB |
| | 99 | 0.78 | 18.2dB | 20.3dB | 25.7dB | 33.7dB |
| | 92 | 0.72 | 17.4dB | 19.2dB | 26.5dB | 34.8dB |
| | 85 | 0.67 | 17.05dB | 18.65dB | 26.85dB | 35.35dB |
| | 78 | 0.61 | 16.8dB | 18.3dB | 27.1dB | 35.7dB |
| | 71 | 0.56 | 16.15dB | 17.7dB | 27.75dB | 36.3dB |
| | 64 | 0.5 | 16.1dB | 17.45dB | 27.8dB | 36.55dB |

TABLE X: Comparative analysis: BCH(127, 113)

| Decoder | At BER of $10^{-5}$ | | |
|---|---|---|---|
| | $E_b/N_0$ | Percentage of Maximum complexity with respect to GRAND | Percentage of Average complexity with respect to GRAND |
| GRAND [53] | 22 dB | 100% | 100% |
| Fading GRAND [60] | 16.2 dB | 969.9321% | 272.3312% |
| ORBGRAND psoft-ZF [61] | 14.1 dB | 47.895% | 32.1709% |
| ORBGRAND psoft-MMSE [61] | 14.1 dB | 59.7907% | 32.1709% |
| ORBGRAND [58] | 13.2 dB | 47.895% | 44.4365% |
| **Proposed DFD** | **21.4 dB** | **0.0146%** | **3.6619%** |

lineates the coding gains facilitated by EDFD. Remarkably,

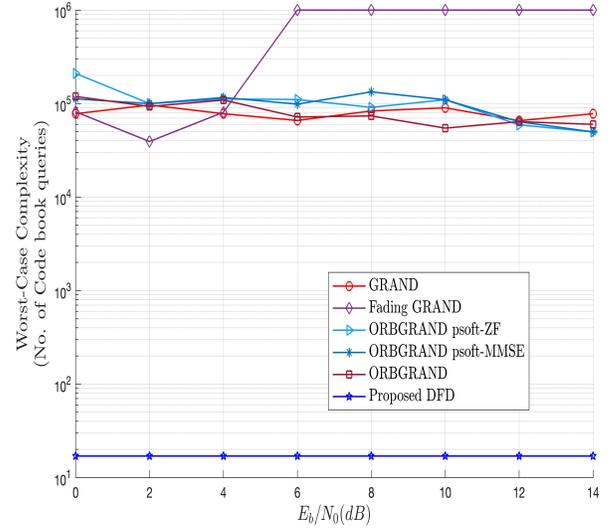

FIGURE 9: Worst-case complexity comparisons in terms of the number of codebook queries of the proposed DFD to existing decoders: $BCH(127, 113)$.

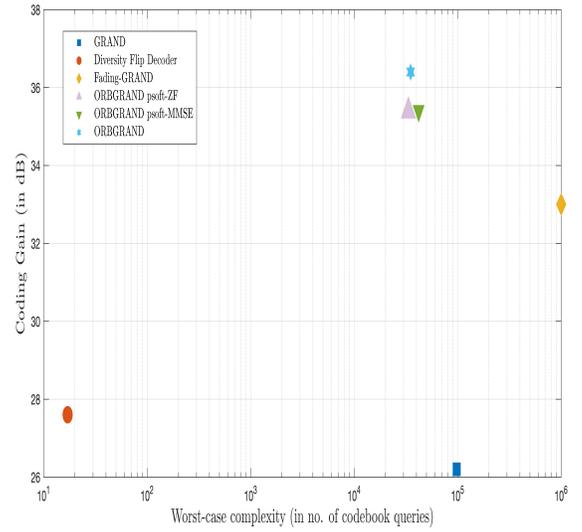

FIGURE 10: Scatter plot of coding gain at BER of $10^{-6}$ vs. worst-case complexity: BCH(127, 113)

observe that as the $\epsilon$ value increases, the BER performance provided by EDFD improves.

In conclusion, the proposed DFD exhibits several key attributes that make it a compelling decoding solution:

- **Universality:** Universally applicable for diverse linear codes.

---

**Algorithm 2:** Extended Diversity Flip decoder

**Input:**
- $\Phi_e$: Flip vector list of EDFD.
- $\underline{\mathbf{r}}$: received codeword.
- $\underline{\mathbf{h}}$: CSI vector.
- $\mathbf{H}$
- $\epsilon$: Extension window length.
- $d_{\min}$
- $n_{\Phi_e}$: Total number of Flip vectors of EDFD.

**Output:**
- $\underline{\mathbf{c}}_o^{\mathrm{opt}}$: The decoded codeword.

1   $\underline{\mathbf{S}}_{\mathbf{y}} = \underline{\mathbf{r}}\mathbf{H}^{\mathbf{T}}$;
2   **if** $(sum(\underline{\mathbf{S}}_{\mathbf{y}}) > 0)$ **then**
3    $[\underline{\mathbf{h}}_{\mathrm{sort}}, \pi_1] = \mathrm{sort}(\underline{\mathbf{h}})$;
4    $\underline{\mathbf{r}}_{\mathrm{sort}} = \pi_1(\underline{\mathbf{r}})$;
5    $\underline{\mathbf{S}}_{\mathbf{y}} = \underline{\mathbf{r}}\mathbf{H}^{\mathbf{T}}$;
6    $le = 1$;     // loop counter initialization
7    **while** $(sum(\underline{\mathbf{S}}_{\mathbf{y}}) > 0)$ & $(le \leq n_{\Phi_e})$ **do**
8     $\underline{\mathbf{fe}}_{\mathbf{pv}} = \Phi_e(le)$;
9     $\underline{\mathbf{r}}_{\mathbf{t}} = \underline{\mathbf{r}}_{\mathrm{sort}} \oplus \underline{\mathbf{fe}}_{\mathbf{pv}}$;
10    $\underline{\mathbf{r}}_{\mathbf{o}} = \pi_1^{-1}(\underline{\mathbf{r}}_{\mathbf{t}})$;
11    $\underline{\mathbf{S}}_{\mathbf{y}} = \underline{\mathbf{r}}_{\mathbf{o}}\mathbf{H}^{\mathbf{T}}$;
12    $le = le + 1$;
13    **end**
14 **else**
15   $\underline{\mathbf{r}}_{\mathbf{o}} = \underline{\mathbf{r}}$
16 **end**
17 **if** $(sum(\underline{\mathbf{S}}_{\mathbf{y}}) == 0)$ **then**
18   $\underline{\mathbf{c}}_{\mathbf{o}}^{\mathrm{opt}} = \underline{\mathbf{r}}_{\mathbf{o}}$;
19 **else**
20   $\underline{\mathbf{c}}_{\mathbf{o}}^{\mathrm{opt}} = \underline{\mathbf{r}}$;
21 **end**

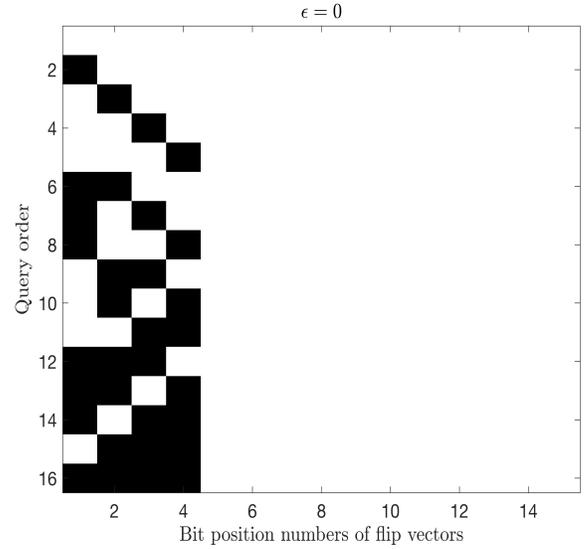

FIGURE 11: Flip vectors pattern of EDFD with $\epsilon = 0$: BCH$(15, 7)$

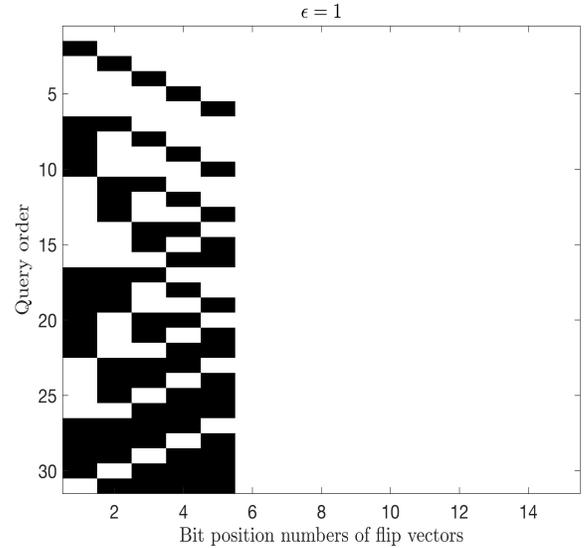

FIGURE 12: Flip vectors pattern of EDFD with $\epsilon = 1$: BCH$(15, 7)$

TABLE XI: Coding gains offered by EDFD over uncorrelated Rayleigh fading channel: BCH(15,7)

| $\epsilon$ | $\mathbf{E_b/N_0}$ at BER of | | Coding Gain at BER of | |
|---|---|---|---|---|
| | $\mathbf{10^{-3}}$ | $\mathbf{10^{-5}}$ | $\mathbf{10^{-3}}$ | $\mathbf{10^{-5}}$ |
| 0 | 11.2dB | 16.6dB | 12.8dB | 27.4dB |
| 1 | 10.15dB | 15dB | 13.85dB | 29dB |
| 2 | 9.3dB | 14.2dB | 14.75dB | 29.8dB |
| 3 | 8.8dB | 13.8dB | 15.2dB | 30.2dB |

- **Preserves diversity in fading channels:** The DFD is capable of preserving diversity order in challenging fading channel conditions.
- **Significantly lower complexity compared to existing approaches:** The proposed DFD imposes a signif-

icantly lower complexity than the existing methods making it ideal for low-complexity applications.

In our future research, we will harness the DFD in diverse dispersive high-Doppler system scenarios of both classical and quantum-coded scenarios.

### Acknowledgment

The authors thank Prof. M. Medard and Prof. K. Duffy for their valuable discussions regarding this paper.

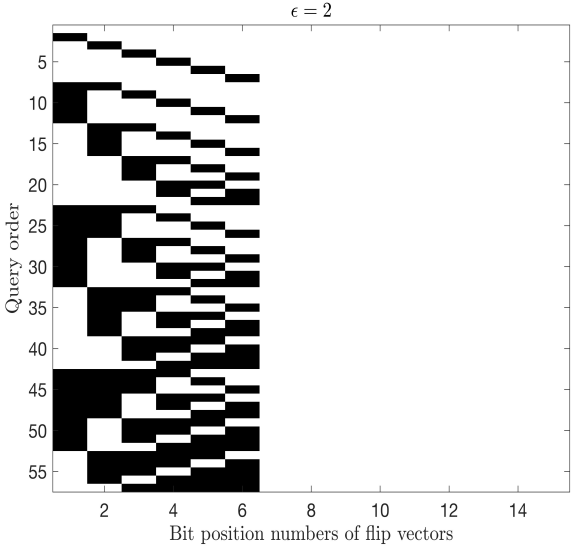

FIGURE 13: Flip vectors pattern of EDFD with $\epsilon = 2$: BCH$(15, 7)$

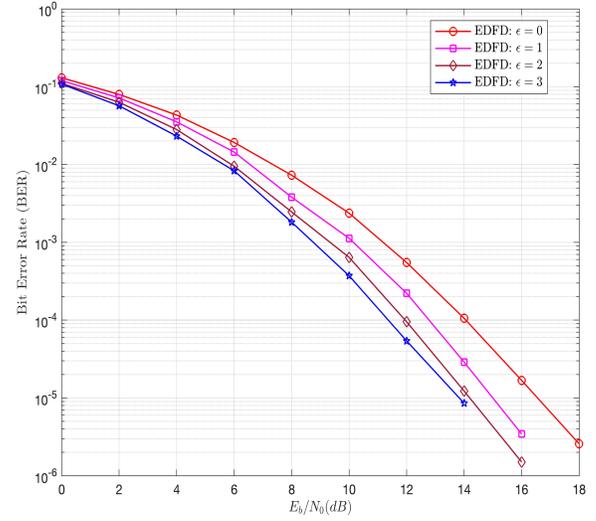

FIGURE 15: BER simulation using EDFD: BCH$(15, 7)$

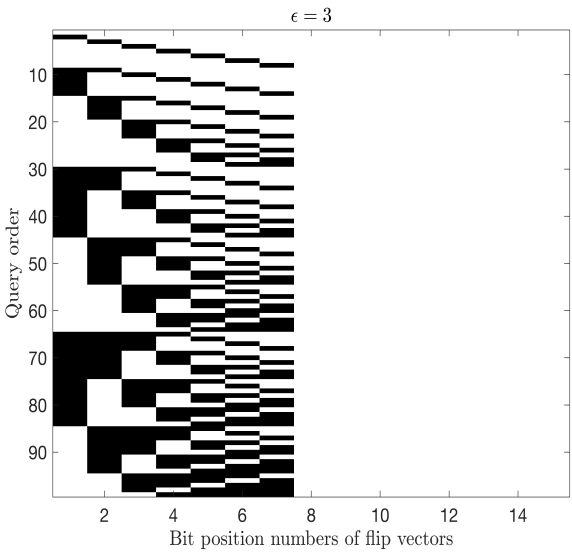

FIGURE 14: Flip vectors pattern of EDFD with $\epsilon = 3$: BCH$(15, 7)$

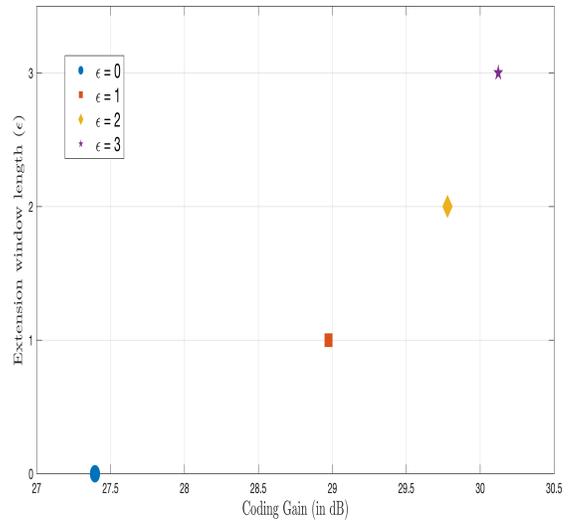

FIGURE 16: Scatter plot of coding gain of EDFD at BER of $10^{-5}$ for different extension window lengths ($\epsilon$): BCH$(15, 7)$

L. Hanzo would like to acknowledge the financial support of the Engineering and Physical Sciences Research Council projects EP/W016605/1, EP/X01228X/1, EP/Y026721/1, EP/W032635/1 and EP/X04047X/1 as well as of the European Research Council's Advanced Fellow Grant QuantCom (Grant No. 789028)

FIGURE 17: Scatter plot of BER (at $E_b/N_0$ of 14dB) versus worst-case complexity (in number of codebook queries) of EDFD for different extension window lengths ($\epsilon$): BCH(15, 7)

TABLE XII: Coding gain of different decoders over uncorrelated Rayleigh fading channel: BCH(127, 113)

| Decoder | $E_b/N_0$ at BER of | | Coding Gain at BER of | |
|---|---|---|---|---|
| | $10^{-5}$ | $10^{-6}$ | $10^{-5}$ | $10^{-6}$ |
| Uncoded | 41.4dB | 51.4dB | 0dB | 0dB |
| GRAND [53] | 22dB | 25.2dB | 19.4dB | 26.2dB |
| Fading GRAND [60] | 16.2dB | 18.4dB | 25.2dB | 33dB |
| ORBGRAND-psoft [61] | 14.1dB | 16dB | 27.3dB | 35.4dB |
| ORBGRAND [58] | 13.2dB | 15dB | 28.2dB | 36.4dB |
| Proposed DFD | 21.4dB | 23.8dB | 20dB | 27.6dB |

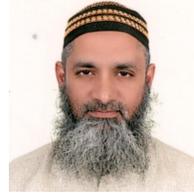

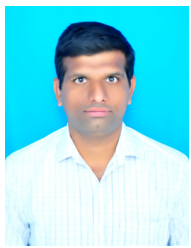

**BERE PRAVEEN SAI** (Student Member, IEEE) received his B.Tech. degree in Electronics and Communication Engineering from the Kakatiya Institute of Technology and Science, Warangal, Telangana, India, in 2013, and his M.Tech. degree in Communications from the National Institute of Technology Warangal, Warangal, India, in 2015. Currently, he is pursuing his Ph.D. at the Indian Institute of Technology Hyderabad, Hyderabad, India. His research interests include coding theory in wireless communications and decoding solutions at short block lengths for 5G and 6G applications.

**MOHAMMED ZAFAR ALI KHAN** (Senior Member, IEEE) received the B.E. degree in electronics and communications from Osmania University, Hyderabad, India, in 1996, the M.Tech. degree in electrical engineering from IIT Delhi, in 1998, and the Ph.D. degree in electrical communication engineering from IISc, Bengaluru, in 2003. He was a Design Engineer with Sasken, Bengaluru, in 1999, a Senior Design Engineer with Silica Semiconductors, Bengaluru, from 2003 to 2005, a Senior Member of Technical Staff with Hellosoft, India, in 2005, and an Assistant Professor with IIIT Hyderabad from 2006 to 2009. He is currently a Professor with IIT Hyderabad. He has more than 15 years of experience in teaching and research. He has made noteworthy contributions to Space time codes. The Space time block codes designed by him have been adopted by the WiMAX Standard. He has been a chief investigator for a number of sponsored and consultancy projects. He is a reviewer for many international and national journals and conferences. He is the author of book Single and Double Symbol Decodable Space-Time Block Codes from Lambert academic press, Germany. His research interests include coded modulation, space- time coding, and signal processing for wireless communications. He was a recipient of INAE Young Engineer Award 2006.

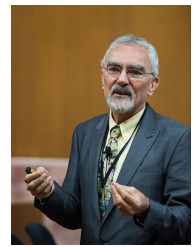

**LAJOS HANZO** is a Fellow of the Royal Academy of Engineering, FIEEE, FIET, Fellow of EURASIP and a Foreign Member of the Hungarian Academy of Sciences. He coauthored 2000+ contributions at IEEE Xplore and 19 Wiley-IEEE Press monographs. He was bestowed upon the IEEE Eric Sumner Technical Field Award.